\documentclass[journal]{new-aiaa}
\usepackage[utf8]{inputenc}
\usepackage{textcomp}

\usepackage{graphicx}
\usepackage{amsmath}
\usepackage{comment}
\usepackage{array}

\usepackage[version=4]{mhchem}
\usepackage{siunitx}
\usepackage{longtable,tabularx}
\title{A mission architecture to reach and operate \\ at the focal region of the solar gravitational lens}

\author{Henry Helvajian\footnote{Principal Scientist, Space Materials Laboratory, AIAA Member. Corresponding author.}, Alan Rosenthal\footnote{Member of Technical Staff, Systems Engineering Division.},
John Poklemba\footnote{Senior Engineer Specialist (retired), Communications Systems \& Engineering Subdivision.},
Thomas A Battista\footnote{Engineering Specialist, Guidance and Control Subdivision, AIAA Member.},\\
Marc D. DiPrinzio\footnote{Senior Project Leader, Systems Analysis and Simulation Subdivision, Senior AIAA Member.},
Jon M. Neff\footnote{Senior Project Leader, CSG Technology.}, and 
John P. McVey\footnote{Associate Director, Systems Analysis \& Simulation Subdivision.}
}
\affil{The Aerospace Corporation, 2310 E. El Segundo Blvd. El Segundo, CA 90245, USA}
\author{Viktor T. Toth}
\affil{Ottawa, Ontario K1N 9H5, Canada}
\author{Slava G. Turyshev\footnote{Research Scientist, Structure of the Universe Research Group.}}
\affil{Jet Propulsion Laboratory, California Institute of Technology, \\ 
4800 Oak Grove Dr., Pasadena, CA 91109, USA}

\begin{document}

\maketitle

\begin{abstract}

We present initial results of an ongoing engineering study on the feasibility of a space mission to the focal region of the solar gravitational lens (SGL).  The mission goal is to conduct exoplanet imaging operations at heliocentric distances in the range $\sim$548--900 astronomical units (AU). Starting at 548~AU from the Sun, light from an exoplanet located behind the Sun is greatly amplified by the SGL.  The objective is to capture this light and use it for multipixel imaging of an exoplanet up to 100 light years distant. Using a meter-class telescope one can produce images of the exoplanet with a surface resolution measured in tens of kilometers and to identify signs of habitability. The data are acquired pixel-by-pixel while moving an imaging spacecraft within the image.   Given the long duration of the mission, decades to 900 AU, we address an architecture for the fastest possible transit time while reducing mission risk and overall cost.  The mission architecture implements solar sailing technologies and in-space aggregation of modularized functional units to form mission capable spacecraft. The study reveals elements of such a challenging mission, but it is nevertheless found to be feasible with technologies that are either extant or in active development.

~

\end{abstract}

\section*{Nomenclature}

\begin{minipage}{0.499\linewidth}
{\renewcommand\arraystretch{1.0}
%%% \noindent\begin{longtable*}{@{}l @{\quad=\quad} l@{}}
\begin{tabular}{@{}l @{\quad=\quad} l@{}}
ACS  &attitude control system\\
ADCS & attitude determination \& control system\\
AU & astronomical unit \\
BOL & beginning of life \\
CDH & command and data handling \\
CBE & current best estimate \\
CEM   & concurrent engineering methodology \\
CTE & coefficient of thermal expansion \\
CONOPS & concept of operations\\
CUC  & critical use case\\
DOE & Department of Energy \\
DSN & NASA Deep space network \\
EP  & electric propulsion \\
EPS  & electric power system\\
FEEP & field emission electric propulsion \\
IoT & internet of things\\
LEO & low Earth orbit \\ 
ly & light year \\
%\\
%%% \end{longtable*}}
\end{tabular}}
\end{minipage}
\begin{minipage}{0.499\linewidth}
{\renewcommand\arraystretch{1.0}
%%% \noindent\begin{longtable*}{@{}l @{\quad=\quad} l@{}}
\begin{tabular}{@{}l @{\quad=\quad} l@{}}
MC & mission capable\\
MMS & mission management software\\
pMC	& proto-mission capable\\
POA & primary optical axis\\
PMM & pulse position modulation \\
PNT & position, navigation and timing\\
RPS	& radioisotope power system\\
RSS	& root sum square\\
Rx & received \\
ROI & region of interest\\
$R_\odot$ & radius of the Sun\\
SEE & single event effects \\
SGL & solar gravitational lens\\
SGLF & SGL focal region \\
SNR	& signal to noise ratio\\
Tx & transmitted\\
TID & total ionizing dose\\
TDM & technology demonstration mission
%\end{longtable*}}
\end{tabular}}
\end{minipage}

\section{Introduction}
\label{sec:intro}

According to Einstein's general theory of relativity, rays of light passing by the Sun are deflected by the angle $\theta= 2r_g/b=1.75 (R_\odot/b)$ arcseconds, where $r_g=2GM_\odot/c^2$ is the solar Schwarzschild radius, $b$ is the trajectory's impact parameter and $R_\odot$ is the solar radius \cite{1-Einstein-1916,2-Einstein:1936,3-Schneider-Ehlers-Falco:1992}. As a result, with its enormous mass, our Sun acts as lens \cite{Turyshev:2017,4-Turyshev-Toth:2017}, focusing the rays of light with the same impact parameter in the focal region that starts at heliocentric distance beyond $z=b^2/2r_g\sim 547.6\,(b/R_\odot)^2$ astronomical units (AU). This is the region where this lens provides enormous brightness amplification by a factor of $4\pi^2 r_g/\lambda=2.11\times 10^{11}\,(1\,\mu{\rm m}/\lambda)$, where $\lambda$ is the observable wavelength, and naturally providing a visual angular resolution of $\lambda/(2R_\odot)= 0.2 \,(\lambda/1\,\mu{\rm m})$ nanoarcseconds  \cite{Turyshev-Toth:2019,7-Turyshev-Toth:2020-extend}.

\begin{figure}[t!]
\centering\includegraphics[width=.703\textwidth]{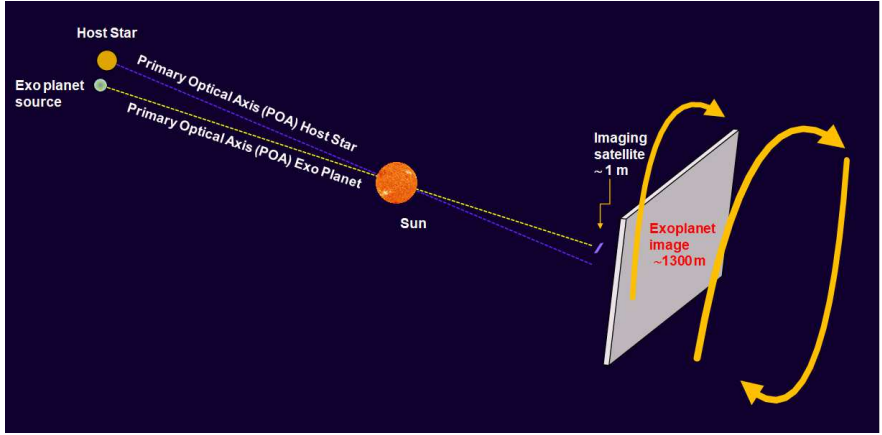}
\caption{
A visualization of the key primary optical (POA) axes and the projected image plane of the exoplanet. The imaging spacecraft is the tiny element in front of the exoplanet image plane.
}
\label{fig1}
\end{figure}

For any heliocentric distance in the focal region, the light field received from a distant object (on the opposite side from the Sun) is focused and amplified by the solar gravity and then received on an image plane that contains a projected image of that object, as shown in Fig.~\ref{fig1}.  This figure also shows the image plane motion.  This results from a combination of solar wobble induced by the gas giants in our solar system and the reflex motion in the host star. The POA of the host star serves as a local navigation reference because its light intensity is orders of magnitude larger than that from the exoplanet. This projection is blurred, it nonetheless represents an image formed by a ``telescope'' of truly gigantic proportions with diameter of that of the Sun \cite{Turyshev-Toth:2019-extend,Turyshev-Toth:2019-blur,Turyshev-Toth:2019-image}. This telescope is called the solar gravitational lens (SGL).

The optical characteristics of the SGL far exceed the capabilities of any present or foreseeable astronomical instrument, leading to the concept of utilizing the SGL itself as an instrument to image faint, distant targets. Of particular interest is the possibility of using the SGL to obtain images of high spatial and spectral resolution of a yet-to-be-identified, potentially life-bearing exoplanet in another solar system in our Galactic neighborhood \cite{9-Turyshev-etal:2020-PhaseII,7-Turyshev-Toth:2020-extend,10-Toth-Turyshev:2020}.

A modest meter-class optical telescope, equipped with an internal coronagraph to block the solar light, operating at the SGL focal region could take advantage of its unique optical properties. While moving away from the sun and by looking back from an instantaneous image plane, light from the distant object appears in the form of an Einstein-ring surrounding the solar disk. The diameter and effective surface area of this Einstein-ring characterize the fundamental optical properties of the SGL: its aperture $\sim2R_\odot$ and light collecting area $\sim4\pi R^2_\odot$. Benefiting from these factors, the telescope may acquire a near-megapixel image, or (trading spatial for spectral resolution) a lower resolution image with spatially resolved spectroscopy, of a well-chosen exoplanet target \cite{7-Turyshev-Toth:2020-extend,10-Toth-Turyshev:2020,Turyshev-Toth:2022-wil_SNR}.

The direct high-resolution images of an exoplanet obtained with the SGL could lead to insight on the on-going biological processes on the target exoplanet and find signs of habitability. By combining spatially resolved imaging with spectrally resolved spectroscopy, scientific questions such as the presence of atmospheric gases and its circulation could be addressed. With sufficient signal-to-noise ratio (SNR) and visible to mid-infrared (IR) sensing \cite{Turyshev-Toth:2022-mono-SNR}, the inspection of weak biosignatures in the form of secondary metabolic molecules like dimethyl-sulfide, isoprene, and solid-state transitions could also be probed in the atmosphere.   Finally, the addition of polarimetry to the spatially and spectrally resolved signals could provide further insight such as atmospheric aerosols, dust, and on the ground, properties of the regolith (i.e., minerals) and bacteria and fauna (i.e., homochirality) (see discussion in \citep{Seager:2010}).

Although the focal region of the SGL begins at large heliocentric distances, successful deep space missions such as Voyager 1/2, Pioneer 10/11 and New Horizons have demonstrated that the capability exists to build spacecraft that can travel to the SGL focal region, operate there successfully, while maintaining reliable communication with the Earth. This opens up a practical possibility of using the SGL to build an astronomical facility with tremendous light gathering power and angular resolution \cite{Friedman-Turyshev:2018,9-Turyshev-etal:2020-PhaseII,Turyshev-etal:2022}.
To accomplish this task, however, it is necessary to deliver a physical instrument to the SGL focal region. To do that one must overcome unprecedented technical challenges that include: i) the enormous distance (four times the distance of our most distant spacecraft to date, Voyager 1), ii) the corresponding very long cruise to the target region, iii) access to reliable communications over this distance, iv) long-term power supply and storage, and v) very precise navigational requirements.

We addressed the challenges above within the context of our ongoing NASA Innovative Advanced Concepts (NIAC) Phase III program \cite{Turyshev-etal:2020-Phase-III} to study a mission concept to the SGL focal region for the purposes of imaging of an exoplanet with spectral and polarization sensitive tools. We have defined a mission architecture which permits a technically feasible mission design to be defined using existing technologies or those in development.  We developed a mission that that relies on solar sailing and small spacecraft (i.e., nanosatellite class) to overcome most of the challenges.  Here we present a mission architecture and concept of operations (CONOPS) for such a mission. The resulting mission design is then used to define technology requirements for future SGL mission studies, tradeoffs, and eventual optimization.

This paper is organized as follows:
Section \ref{sec:science} presents the primary science requirements and mission drivers for an imaging mission to the SGL focal region.
In Section~\ref{sec:tech_mission} we present a technical overview of the SGL mission, experiment and methodology used to design the mission. Section \ref{sec:miss-arch} discusses one of the possibilities for an SGL mission architecture that relies on nanosatellites propelled by solar sailing to high solar system exit velocities. We provide an overview of the mission design process, relevant trade-offs and discuss several key mission drivers.
Section~\ref{sec:tech-road} presents some of the technologies that are may expedite the development of an SGL mission.
In Section~\ref{sec:concl}, we conclude with a discussion of the critical technology areas which need further development.

\section{Scientific requirements and mission drivers}
\label{sec:science}

The SGL offers capabilities that are unmatched by any planned or conceivable optical instrument, including large  telescopes, optical phased arrays or synthetic aperture very-long-baseline optical interferometry. With its unique optical properties, the SGL can be used to obtain detailed, high resolution images of Earth-like exoplanets as far as  100 light years (ly) from Earth, with measurement durations lasting months, or at most a few years.

A mission would begin with the selection of the intended target, an Earth-like exoplanet orbiting within the habitable zone of its host-star and likely showing the presence of an atmosphere or other signs of life-bearing conditions.

\subsection{Key scientific objectives}

The SGL imaging mission is designed to deliver high spatial-resolution images with spectroscopy of an Earth-like, potentially habitable exoplanet in our stellar neighborhood. Its main limitation is the requisite data collection (integration) time, which is measured in months or years. Predictable short-term, transient phenomena, such as diurnal changes, must be estimated and accounted for in the image reconstruction. Unpredictable short-term phenomena will be treated as noise that is mitigated and removed by the long integration times. Even with these limitations, the SGLF mission can produce data of unmatched scientific value.

The SGL mission can offer resolved imaging of an exoplanet, search for biosignatures, and study temporally varying properties of the target. These primary investigation areas are briefly discussed below.

\subsubsection{Resolved imaging of primary surface features}

The mission will produce two dimensional (2D) surface topographic maps that can identify global landforms (e.g., continents, mountains, polar ice caps, etc.)  The mission will be able to measure spatially resolved structures and primary topographical features on the surface of a target exoplanet. The imaged data  will permit inferences to be made on current geological conditions that drive the potential biological processes. The mission could document  seasonal global patterns. While other techniques could at best provide indirect measurements of global planetary  averages, the SGL mission offers direct data on a spatially resolved scale to characterize the surface distribution of sources correlating with continental maps and topography. This also applies to technological signatures that  may be present at a detectable level (for instance, if present, artificial light from large population centers might be detectable when those locations are on the dark side.)

\subsubsection{Search for possible biosignatures}

The mission will use both spectrally resolved data (e.g., looking for the presence of atmospheric gases) and spatially resolved data to search for clues that could signify on going bio-processes to indicate habitability. It will provide data to study atmospheric circulation. There are a wide range of trace atmospheric constituents that are not well studied because their features are too weak with modern Earth based facilities. The mission will detect these features and will be able to detect a habitable world unambiguously.

\subsubsection{Study of temporally varying phenomena}

The mission should be able to measure periodic temporal phenomena, e.g., planetary rotation, seasonal variations, diurnal variations, large-scale weather patterns. It will detect signatures of stochastic processes such as cloud  formation. It will study variability of the on-going planetary processes and their input into life-support conditions. The SGL mission is driven by a set of key science objectives that, in turn, imply a set of technical requirements.

These requirements can be succinctly summarized as: The ability to obtain images of high spatial and/or spectral  resolution of a selected exoplanet with a mission duration lasting  $\lesssim50$ years including cruise and science operations.

\subsection{Basic mission drivers and trade-offs}

The science and mission requirements outlined in the preceding section translate directly into a set of fundamental mission drivers that an SGL mission must meet in order for it to be successful. For the purposes of this analysis, we assume an Earth-like exoplanet at $z_0=100$~ly, at $  z=650$~AU from the Sun which projects an image with diameter  of $d_\oplus=2R_\oplus  (z/z_0)\simeq 1,300\,(z/650~{\rm AU})(100~{\rm ly}/z_0)$~meters  that becomes progressively larger at greater heliocentric distances.  Table I summarizes the science and technical requirements of an SGL mission.

\begin{table}
\caption{\label{tb:req}Fundamental science and technical requirements of an SGL mission.}
\begin{center}
\vskip -10pt
\begin{tabular}{|l|c|}
\hline
Description&Value\\\hline\hline
Distance from the Sun, $  z$& $\sim$650--900~AU\\
Cruise time&$<30$~years\\
Science operations&$\sim 10$~years\\
Local reference frame position error&$<1$~m\\
Telescope aperture, $d$&$>1$~m\\
Telescope field-of-view&$\gtrsim 3.5''$\\
Coronagraph performance&$<10^{-7}$\\
Telescope bit depth&$\gtrsim 24$~bits\\
Sampling frequency&$\lesssim 3$~min\\
Sustained COMM downlink rate&$\gtrsim {\cal O}(100)$~bps\\
Cumulative $\Delta v$: Science operations&$\gtrsim 1890$~m/s\\
Image acceleration&$\sim 6\times 10^{-6}$~m/s$^2$\\
\hline
\end{tabular}
\end{center}
\end{table}

\subsubsection{Operating heliocentric distance}

Although the SGL focal region begins at 547.6 AU from the Sun, at this distance, the Einstein ring formed by a distant source would appear to touch the solar disk. It is only at $ z\sim 650$~AU that the separation between the Einstein ring and the solar disk becomes large enough at visible wavelengths for the two to be distinguishable by a meter class telescope. Operating at greater distance from the Sun offers significant benefits, as the increasing separation between the Einstein ring and the Sun implies better coronagraph performance, less light contamination from the Sun, and also a reduced corona background.

\subsubsection{Cruise duration}

Reaching the SGL focal region at 650 AU in less than 25 years implies a hyperbolic escape velocity in excess of 25 AU/year. We considered current and projected performances of several known propulsion techniques that are either operational or in development, including chemical propulsion (3--4 AU/yr (i.e., 15--20 km/sec) velocities (e.g., NASA New Horizon $\sim$16 km/s), solar thermal, nuclear thermal and electric ($\sim$10--12 AU/yr) [9], solar sails and electric sails.    Of these, solar sailing in combination with a close ($\lesssim 0.1$~AU) perihelion represents the only method of propulsion that is sufficiently mature for a realistic mission to reach solar system exit velocities of 20+ AU/year. Solar sails, however, require a very large surface area to mass ratio, which implies small, lightweight spacecraft, i.e., a smallsat/nanosatellite with spacecraft mass $\lesssim$100 kg.

\subsubsection{Science mission phase duration}

Once it reaches the SGL focal region, the mission can continue to egress the solar system as it makes observations.  Given the long integration times required in order to achieve the necessary SNR, the mission must remain operational for at least 10 years of near continuous science observations. The observations need not be of one exoplanet target but because of optical scaling, it could be the entire exosolar system.

\subsubsection{Telescope aperture}

The SGL mission requires an optical instrument that can, from a distance greater than 650 AU, image the Sun and its  immediate  neighborhood at sufficient  resolution to  distinguish  the  Sun  from the  Einstein ring.    The angular separation between the Einstein ring and the solar disk, viewed from a distance z, is calculated as $\delta\theta = \sqrt{2r_g/z} - R_\odot/z$, where $r_g=2GM_\odot/c^2$ is the solar Schwarzschild radius. The angular resolution of a diffraction-limited telescope is given by $\theta\sim 1.22\lambda/d$, where $d$ is the aperture and $\lambda$ is wavelength.  For visible and near IR wavelengths coupled with simulations that treat the solar corona as a noise source, a requirement emerges that the telescope has to be meter-class to effectively utilize the SGL for visible imaging. Moreover, the telescope aperture and its resolution also places constraints on the image sensor (i.e., number pixels).  For example, a 1~m telescope and 500 nm wavelength has a resolution of 0.126''.  The angular diameter of the Sun at 650~AU is $\approx2.9$ arcseconds ($''$), yielding a circumference of $9.3''$.  Parsing this circumference into telescope resolution elements yields 74 resolved pixels.

\subsubsection{Coronagraph}

Signal photons arrive from the Einstein ring that is formed circumferentially by the SGL around the solar disk. The telescope system must be able to image the Einstein ring surrounding the Sun, which implies a field of view of $\sim 3.5''$, decreasing over time as the distance between the instrument and the Sun increases. In the optical bandwidth, light from the solar disk must be blocked at least to a factor of $10^{-7}$ by a coronagraph to minimize light contamination.

\subsubsection{Navigational accuracy}

Successful imaging requires precise knowledge of the telescope position within the projected exoplanet image in the image plane. This implies a requirement to navigate to the image region, find, and reliably identify the location where the exoplanet image appears, and then determine the instrument position with respect  to this image with meter-scale accuracy. Because the dynamics of the moving image need to be well understood, establishing a local inertial reference frame e.g., by using auxiliary spacecraft helps.  One measures the motion of the observing spacecraft with respect to this inertial frame.

\subsubsection{Flight dynamics}

The noninertial motion of the projected image of an exoplanet requires continuous or periodic velocity adjustments for the imaging telescope.  The total change in velocity during a 10-year science mission depends on the specific target. We calculate a $\sum\Delta v\sim 1,890$ m/s for a typical Earth-like exoplanet for the simulated image acceleration of $\sim 6\times 10^{-6}$~m/s$^2$ (see discussion at \cite{14-Turyshev-Toth:2022-wobbles}).

\subsubsection{Communications}

There are two components that drive the communication requirements:  1) maintaining spacecraft operations during the whole mission and 2) transferring the data from the science phase to Earth.   The former is based on the extent of autonomy present on the vehicle which defines the architecture of the mission management software, while the latter depends on the on-board processing capabilities for image reconstruction and the trade-offs between spatial and spectral resolution.  It may be necessary to downlink the raw data. The science data rate will be determined by sensor (spatial and spectral) resolution and sampling frequency, along with any compression algorithms that may be used for the downlink, but the communications equipment on board must be capable of sustained high speed communication from 650--900 AU during the science phase.  The list of requirements that can be satisfied by processing the raw science observations on the ground include a) the removal of the solar corona background, b) light contamination from other sources including the host star, and c) the need to deconvolve the image that is blurred by the SGL PSF.

\subsection{Technology challenges}

Having established the key scientific objectives and the basic mission drivers necessary to meet the objectives, we assessed the basic technology challenges.

\subsubsection{Propulsion}

As the first step in this process, we focused our attention on propulsion technology. Despite the impressive progress made in chemical propulsion with the introduction of a new generation of heavy-lift launch vehicles (i.e., Falcon Heavy (SpaceX), Space Launch System (NASA) and others), our earlier conclusions \cite{9-Turyshev-etal:2020-PhaseII} stand: A direct ascent approach, using chemical propulsion cannot deliver a midsize telescope to heliocentric ranges beyond $\sim$550 AU within the desirable 25--30 years transit times. Nuclear fission based propulsion may work as an alternative, but currently it is prohibitively expensive.

The ongoing development of NEA Scout (NASA), recent selection of Solar Cruiser (NASA), development of several   solar sailing spacecraft in ESA and JAXA, demonstrates the increased interest in solar sailing technology. In addition, new sail materials with improved performance are being developed. These materials can be used to manufacture sails that are very thin ($<1~\mu$m), strong, lightweight and have high reflectance. The use of these materials will allow for the close solar perihelion passages ($\sim$10--15 $R_\odot$)  that are needed to reach velocities of 20-25 AU per year while surviving the thermal loads associated with such close proximity to the Sun. It is anticipated that sail surface areas will be large.  For example, a 20 kg mass sailcraft with a perihelion pass  $\sim$10--15 $R_\odot$ requires a sail area of $3000~{\rm m}^2$.

Using the data collected by the Parker Solar Probe \cite{Venzmer-Bothmer:2018}, we examined the relevant thermal and dynamical environments in the immediate solar vicinity (i.e., 5--25 $R_\odot$).  The conclusion is that the expected thermal and mechanical loads are consistent with our early design assumptions.  Leaving aside the potential for dust particle impacts \cite{Vaverka-2021} temperature changes along the intended trajectory of the SGL spacecraft are predictable, reaching the peak value of 700${}^\circ$C. The relevant dynamical environment is benign. In general, the stresses on the sailcraft structure can be well understood. For the sailcraft, we considered among other known solar sail designs, one with articulated vanes (i.e., SunVane). While, currently at a low  technology readiness level (TRL), the SunVane  does permit precision trajectory insertion during the autonomous passage through solar perigee.  In addition, the technology permits trimming of the trajectory injection errors while still close to the Sun.  This enables the precision placement   of the SGL spacecraft on its path towards the image cylinder which is 1.3 km in diameter and some 600+ AU distant.

\subsubsection{On-board power}

On-board power presents one of the more significant challenges for the SGL mission. The SGL mission needs lightweight, long-life nuclear power sources. Moreover, it would benefit if the power sources were small (e.g., 1--10~W electric power), allowing them to be distributed where needed. Recently, there has been some progress in this area, motivated by the National Academy of Science (NAS) Decadal Surveys in Planetary Sciences and the emerging demand from some branches of the US Federal Government. One particular example is a conceptual power system developed at JPL with a mass of $< 5$ kg with a half-life over 90 years \cite{Abelson:2004}. A second example is a project sponsored by NASA/STMD under NASA's Innovative Advanced Concepts (NIAC) program called APPLE \cite{22-Nemanick-etal-2022}.  These and efforts at the Oak Ridge National Laboratory may help resolve this problem.

\subsubsection{Communication}

The SGL mission may need to have two forms of communication:  1) Optical communications for the data return to Earth and 2) low power RF communications to establish a local area network (LAN) among mission spacecraft.  The latter is handled by near omnidirectional patch antennas while the former is integrated as part of the mission payload.  Link budget studies show that RF communication from 900~AU is not feasible but possible with optical communications.  Laser communications does raise concerns over the reliability, longevity and performance degradation of a laser transmitter over a very long-duration mission.  However, we have noted a significant increase in the reliability of lasers in the past five years alone along with an increase in maturity of laser communications in space systems \cite{Talamante-2021}.  Moreover, NASA just completed a 10-year technology demonstration project (Laser Communications Relay and Demonstration\footnote{Laser Communications Relay and Demonstration (LCRD): https://www.nasa.gov/mission\_pages/tdm/lcrd/index.html} (LCRD)) lead by NASA/Goddard with a goal to move TRL from 6 to 8.

\subsection{Approaches to meeting the technical challenges}

Formulating the scientific objectives and the mission drivers allowed us to identify key technology challenges.  In this section we present technical approaches against the challenges.

\subsubsection{Telescope aperture}

Based on our simulations, a minimum telescope aperture of 1 meter is required and a larger aperture would be preferred (e.g., 2-5~m) but that changes the primary mirror design from that of a single optical structure to one that is segmented or an interferometric telescope.  Segmented optics produces high spatial frequencies that the coronagraph must remove while the interferometric approach adds complexity.  The benefit of larger aperture is the increase in the per-pixel SNR during data collection and reduction of the integration time by a factor of $1/d^3$, where $d$ is primary mirror diameter.

We explored the feasibility of using foldable, ``stretchable'', or sparse aperture or segmented optics to increase the aperture size larger than 1~m. In all these options, additional  technology hardware needs to be included in the space vehicle for optical wavefront sensing and control of the deployed optic. While instruments for wavefront sensing may have low mass, currently it is the control systems, needed for the adaptive optics, that tend to be heavy.  There are technology developments and some commercially available components already that permit the manufacturing  of segmented mirror designs which are integrated with adaptive optics. This is a technology that is advancing and could be of benefit to the SGL mission.

\subsubsection{In-flight assembly}

A 1-meter telescope, integrated with the necessary coronagraph, communication, avionics, batteries, and other subsystems cannot currently be produced in a 10-20 kg mass and within a volume of a 6U+ CubeSat. Our initial analysis shows that in totality, an MC spacecraft would likely have a mass of $\sim$100 kg, which cannot be accelerated to the desirable 20 AU/year velocities without excessively large solar sails.  While materials science develops structural materials for sailcraft (e.g., booms) that could propel a 100 kg payload to $> 20$ AU/yr velocities,  we consider the use of in-flight robotic assembly of modular segments delivered on multiple, fast-moving sailcraft accelerated by solar radiation pressure.  The architecture relies on the fast-maturing technology of nanosatellites (10-20 kg) and permitting flight implementation with solar sails in less than a decade \cite{9-Turyshev-etal:2020-PhaseII}.

\subsubsection{Sail technology with articulated vanes}

A TDM has been studied to test the solar sail concept developed by and patented by L'Garde and NXTRAC. The current baseline TDM \cite{Garber-etal:2022} vehicle consists of six  20 m${}^2$ vanes attached to a 6-m hexagonal carbon truss. The mass of the entire vehicle is 5.8 kg with a resulting area-to-mass ratio of $A/m \sim23~{\rm m}^2$/kg, a factor of 2 greater than any current or planned planar sail missions but below the $\sim 150~{\rm m}^2$/kg ratio necessary for the SGL mission.

\subsubsection{Sail Materials, shapes and controls}

Our current focus for sail material is metal coated graphene films.  Fabrication of such films is presently in progress.  For example, aluminized graphene can reach perihelion down 0.1~AU ($25R_\odot$, and potentially lower with proper thermal emissivity designs) without melting. The SGL mission requirements are perihelion at 10--15$R_\odot$, where the temperatures can reach 700~K (15$R_\odot$, 95\% reflective, emissivity, $\epsilon=0.8$). The estimated areal density of this material combination, 1.5~g/m${}^2$, is quite appealing. TRL of current sailcraft components is:
i) Square sail shape have TRL 8--9. Such sails have been flown and tested. Solar Cruiser with 1,600 m${}^2$ will fly in 2024.
ii) Solar sail deployment technology: TRL 9. Tested on multiple prior missions. iii) Sail attitude control controls \cite{Sperber-Eke:2016}: Reflectance modulation (TRL 9, IKAROS), mass translation (TRL 7, NEA Scout), vanes (TRL 5-6, Sunjammer).
iv) Materials: TRL 2 -- theoretical predictions have been made, fabrication and experiments are in progress \cite{Davoyan:2021}.

Several sail control approaches have been evaluated, including reflective patches (similar to those on Solar Cruiser), reflective flaps (similar to those on Sunjammer), translation  of mass (similar to that on NEA Scout). We investigated reflective solar patches in which the material reflectivity is altered and the use of movable reflective flaps. The data shows the rate of change for an 85-degree sail rotation (10 kg payload) are possible at rates up to 1 deg/s.   These results suggest that reflective flaps are efficient at turning a sail but at rotation rates up to 1 deg/s.  For the SGL mission, much smaller turn rates are needed, which can be implemented.

\subsubsection{On-board power systems}

An initial design of the SGL architecture estimated $\approx100-150$ W of power will be necessary for the mission per SGL sailcraft.  A trade study was done for various radioisotope power sources (RPS) that would meet the requirements.  NASA's eMMRTG has beginning-of-life (BOL) of 160 W (electric) but is too massive (44~kg).  The advanced Stirling RTG (ASRG) can deliver 130 W but in a 32 kg package.   The baseline design for the SGL mission is a \textonehalf-scale Brayton-type RPS (BOL 55 W, $\sim7$ kg) but in multiple units.  We chose the Brayton design to gain an increase in efficiency over traditional radioisotope thermoelectric generators (RTGs) but at a cost of a moving part that could potentially fail.   The reliability of Brayton RPS have been shown to approach 20 years.

\section{\label{sec:tech_mission}Technical overview of SGL mission, experiment, and methodology}

For an Earth-size exoplanet 100 ly behind the Sun, when the image of that exoplanet is projected by the SGL it forms an image $\sim1.3$ km in diameter at 650 AU.  This analysis is simple geometric optics.  A meter-class telescope, looking back at the Sun, cannot image the whole exoplanet but by blocking the Sun will observe an Einstein ring which represents a single image pixel.  The information in that image pixel can be traced back to a location on the exoplanet.  However and because the SGL has spherical aberration, the captured light is a blend of light rays from the whole planet. Moreover, the light rays from the directly imaged region are a small part of the total collected photon flux. The result is a blurred projection in the image plane  \cite{7-Turyshev-Toth:2020-extend,10-Toth-Turyshev:2020}.

\subsection{\label{sec:miss_descript}Mission description}

The goal of a SGL imaging mission is to collect light from an image pixel at a time and use it to reconstruct images of the source with a variety of spatial or spectral resolutions. A single spacecraft could do this, but to achieve the required navigational accuracy, mitigate risk of failure and increase data acquisition rate, the SGL mission is best accomplished  by delivering and operating a group of identical, fully functional spacecraft to the SGL focal region.

A prerequisite of this mission is that the Sun must be blocked with high precision.  There are at least two primary approaches.  Using an internal coronagraph or an external occulter (i.e., a starshade).  The former reduces mission complexity at the expense of increased light scattering internal to the telescope resulting in less overall signal, while the latter adds complexity to the mission by requiring two spacecraft to fly in sequent formation, precisely, but with the result of increase in the SNR.  Simulations of the SGL data acquisition show that at least a meter class telescope is necessary for practical SNR for a coronagraph based telescope while a 0.40~m aperture is sufficient given an occulter \cite{Turyshev-Toth:2022-mono-SNR}.  Moreover, the occulter based system permits measuring wavelengths in the medium to long wavelength IR.

 In this study, we focused on reducing mission complexity by using a coronagraph (i.e., of Lyot type) as the baseline.    Consequently, each spacecraft carries an optical telescope with 1-m aperture, equipped with a coronagraph capable of sunlight rejection at the level of $10^{-7}$  or better. To ensure that the collected image pixels can be deconvolved to remove spherical aberration, the platoon of flying vehicles must follow the motion of the exoplanet's  image as projected by the SGL in the image plane with the precision of 0.1 m. The process entails imaging the Einstein ring, at various locations within the image plane and sending the raw data back to Earth.

To satisfy constraints of on-board consumables, the SGL mission must reach the SGL focal region in $<30$ years from launch. Successful science operations require the spacecraft to be far enough from the Sun for there to be sufficient separation between the Einstein-ring and the solar disk. For this, the spacecraft needs to reach $\gtrsim650$~AU from the Sun \cite{7-Turyshev-Toth:2020-extend,10-Toth-Turyshev:2020}.

\subsection{Solar sail propulsion}

\begin{figure}
\hskip -14pt
\includegraphics[scale=0.63]{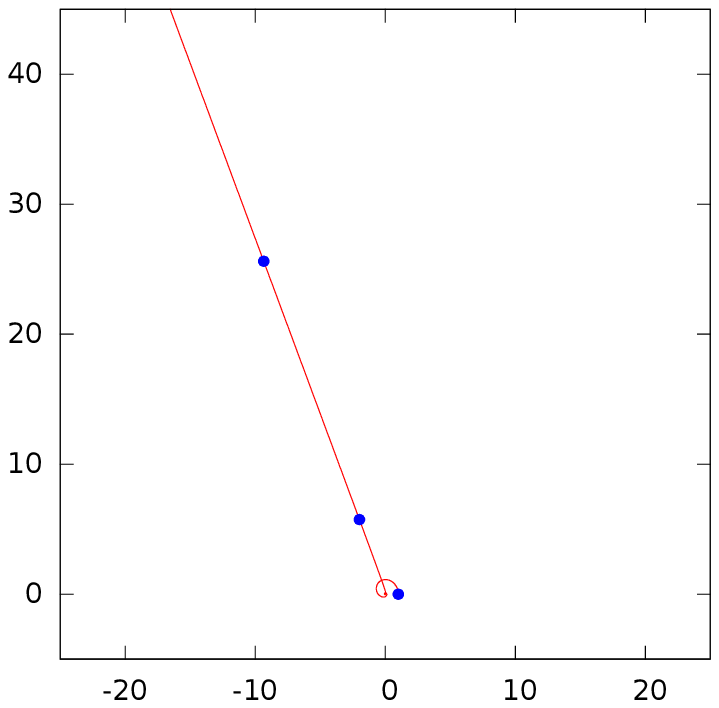} \hskip -13pt
\includegraphics[scale=0.63]{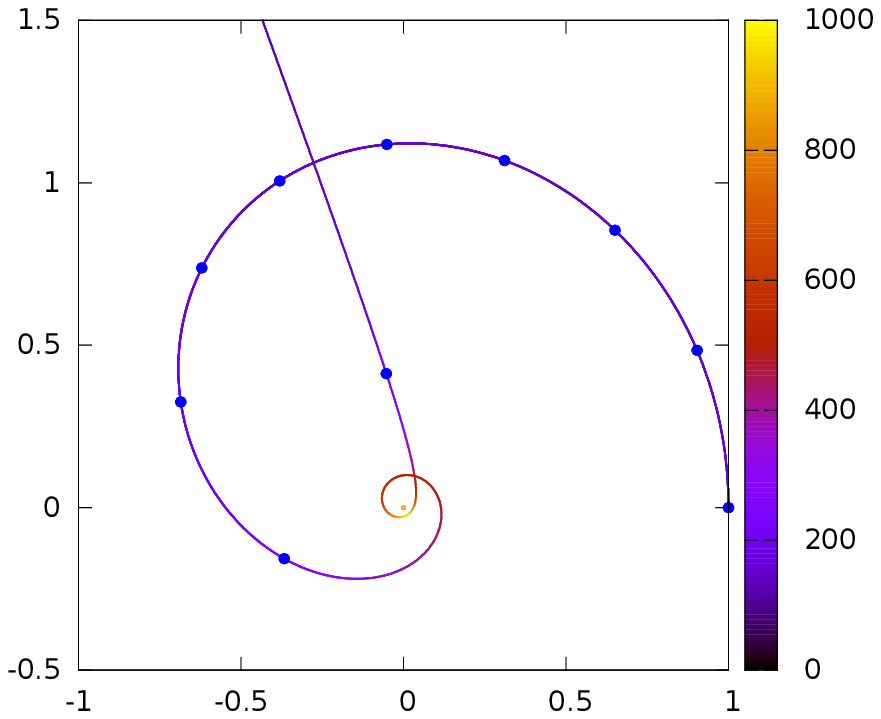} \hskip -13pt
\includegraphics[scale=0.63]{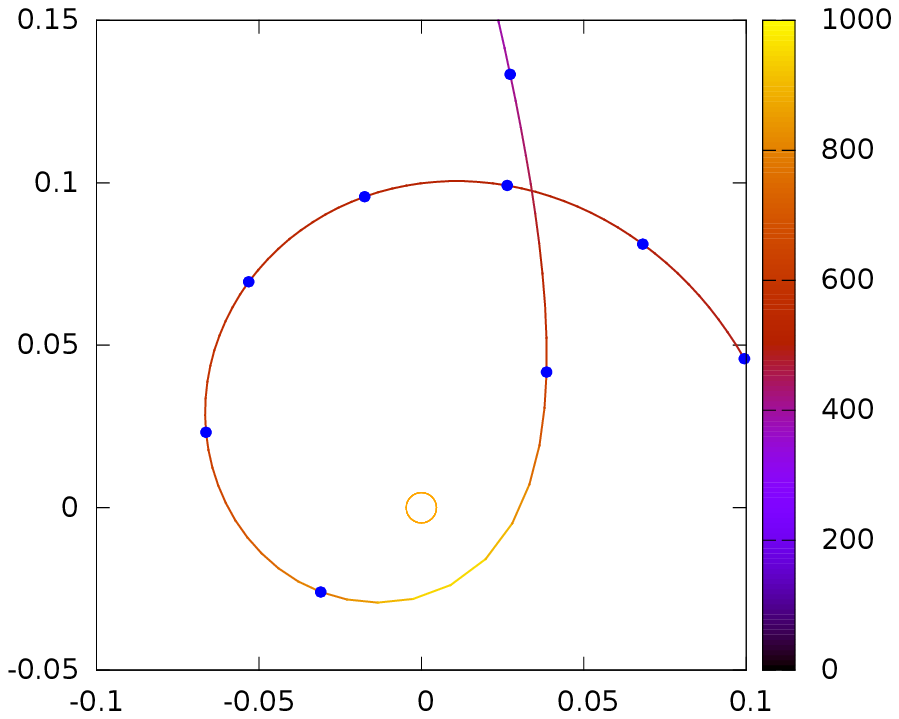}
\caption{\label{fig:sails}
A representative (unoptimized) sample solar sail trajectory for a $10^4~{\rm m}^2$ sail, 50~kg spacecraft achieving $21$~AU/year with a perihelion of 0.025~AU. Both axes are in AU. Left: the first 50~AU; markers represent years. Middle: The inward trajectory; markers at months. Right: The innermost part of the trajectory; markers represent days. Color scale is in K.}
\end{figure}

Solar sails use solar radiation pressure for thrust and subsequently do not require a reaction mass and are therefore not constrained by the limitations of the rocket equation. As several previous studies suggest (\cite{Friedman-Garber:2014} and references therein), a sailcraft can be propelled by a sufficiently low perihelion pass to be placed on hyperbolic trajectories with very high excess velocities ($v_{\rm inf}$). The hyperbolic velocity after solar perihelion depends on the perihelion distance, the sail area ($A$) and the total mass ($m$) of the sailcraft. The bound on the maximum cruise velocity that a solar sail may reach is given by the equation \cite{Davoyan:2021}
$$
v_{\rm inf}\simeq \sqrt{2(2\rho+\alpha)\frac{S_{\rm 1AU}}{d_{\rm per}}\frac{A}{m}-\frac{2 G M_\odot}{1{\rm AU}+d_{\rm per}}},
$$
where $G$ is the gravitational constant,
$M_\odot$ is the solar mass, and $S_{\rm 1AU}$ is the solar irradiance at 1 AU. Also $\rho$ and $\alpha$ are the optical properties of the sail (i.e., solar reflectivity and absorptivity), $A/m$ is the sail area to total sailcraft mass ratio, and $d_{\rm per}$ is the solar perihelion distance.

A closer perihelion approach reduces the $A/m$ ratio, but it is the design of the sailcraft and its material properties that define the closeness of the approach. The sail material has to be highly reflective and minimally absorptive on the sun facing-side while having high emissivity on the back side. Therefore, it is a thermal balance problem between absorptivity and emissivity of the composite materials that make up the sail and support booms.

A representative (unoptimized) example trajectory is shown in Fig.~\ref{fig:sails} using a 50~kg sailcraft with a $10^4~{\rm m}^2$, achieving $21$~AU/year with a perihelion of 0.025~AU, at maximum temperature of 950~K for sail material with reflectivity of $\rho=0.95$ and emissivity of $\epsilon=0.80$. An optimized trajectory may achieve the same or higher egress velocity using a larger perihelion and a lower maximum temperature.

Table~\ref{tab2} lists the estimated temperatures for a composite material with various properties. It shows that polymeric materials (e.g., CP1, polyimide, Kapton) will not do well, while a thin Al metal film is a good reflector ($\sim$93\% reflective, melts 660${}^\circ$C), it is too heavy for large sail areas. A 1~$\mu$m free-standing Al film over an area of 1500 m${}^2$ has 4 kg mass.   Thin ceramics (e.g., TiO${}_2$, TiN) and metamaterials could work if the manufacturing could be scaled to large areas.  For $A/m$ $=150~{\rm m}^2$/kg and a 10 kg vehicle (assuming a square sail), suggests a sail size $\sim39$ m on a side.  NASA's Solar Cruiser  destined to Sun-Earth L1 mission (see \cite{17-Pezent-etal:2019}, est. launch 2025) has a sail area $\sim41$ m on side but uses CP1 polymeric material.    Confining the sails ($\sim$1500--2000~${\rm m}^2$), the structurally stiff booms and a storage container to a few kg, suggests a material ``system'' with areal mass density $<2.5~{\rm g/m}^2$.

\begin{table}[h!]
\begin{center}
\caption{\label{tab2} Estimated temperatures for a composite material at perihelion of 10--15$R_\odot$. Notations used $\rho, \alpha$ and $\epsilon$ stand for reflectivity, absorptivity, and emissivity, correspondingly.}
\begin{tabular}{|p{0.36\textwidth}|p{0.16\textwidth}|p{0.16\textwidth}|}
\hline
 &\multicolumn{2}{c|}{Temperature at perihelion}\\\cline{2-3}
Composite Material & perihelion at 15$R_\odot$ &perihelion at 10$R_\odot$ \\\hline\hline
$\rho=0.90$ ($\alpha=0.10$), $\epsilon = 0.8$ (e.g., asphalt)   & ~~~ 810 K (538${}^\circ$C) &  ~~~  960 K (687${}^\circ$C)            \\
$\rho= 0.95$ ($\alpha=0.05$), $\epsilon = 0.8$ (e.g., asphalt)   & ~~~ 682 K (409${}^\circ$C) &  ~~~  807 K (534${}^\circ$C)          \\
\hline
\end{tabular}
\end{center}
\vskip 0pt
\end{table}

A sailcraft with articulated vanes allows the control of the egress thrust vector. For the SGL mission, the control system needs to allow for a perihelion passage with the following state requirements: Position knowledge 1 km RSS, velocity knowledge 1~cm/sec, and attitude knowledge 1 deg and 1 deg/sec rate (all 1-$\sigma$ values)\cite{Garber:2021}.    A number of thrust control sailcraft designs have been considered (e.g., reflectance modulation-IKAROS \cite{16-Tsuda-etal:2013}, mass translation-NEA Scout \cite{17-Pezent-etal:2019}, vanes-Sunjammer \cite{18-Barnes-etal:2014}) along with shapes (circle, square, vaned).  Investigations are underway to determine the configuration and control scheme that permits the highest fidelity control of the thrust vector.

\subsection{\label{sec:4.1}Spacecraft description}

An SGL mission comprises a constellation of identical spacecraft, required for redundancy, precision navigation, removal of background light contamination and optimized data return.

%%%%%%%%%%%%%%%%%%%%%%%%%%%%%%
\begin{figure*}[h!]
  \vspace{-7pt}
  \begin{center}
\includegraphics[width=0.254\textwidth]{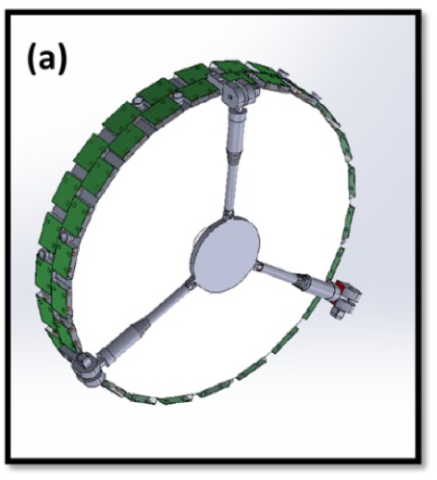}
\includegraphics[width=0.304\textwidth]{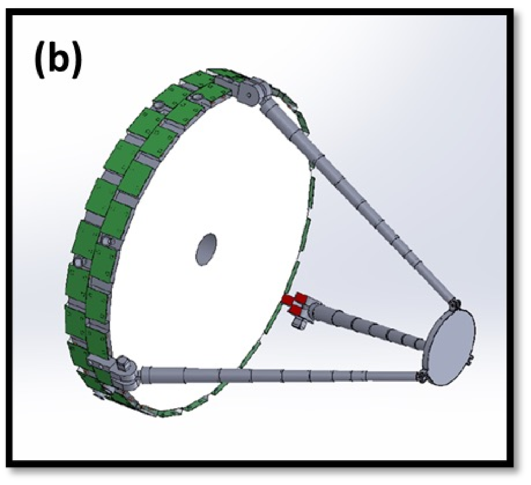}
  \end{center}
  \vspace{-5pt}
  \caption{(a) SGL spacecraft with secondary optical mirror stowed and (b) deployed. Both are shown with tiled RPS technology. The mirror is 1~m in diameter and the thickness of the space vehicle body is $\sim$20~cm.
  }
\label{fig:sc-concept}
\end{figure*}
%%%%%%%%%%%%%%%%%%%%%%%%%%%%%%

\begin{table}[h!]
\begin{center}
\caption{\label{tab3a} Primary components of notional SGL vehicle.}
\begin{tabular}{|c|l|}
\hline
\#& Vehicle component \\
\hline\hline
1 & 6U CubeSat bus (avionics)\\
2 & Solar sail and holder\\
3 & Primary mirror -- telescope\\
4 & Optical COMM system\\
5 & Science package (telescope, coronagraph)\\
6 & Boom arms with secondary mirror (telescope)\\
7 & Batteries\\
8 & Radioisotope power system (RPS)\\
9 & Reaction wheels\\
10 & Electric propulsion\\
\hline
\end{tabular}
\end{center}
\vskip 0pt
\end{table}

The main components of a single nanosatellite (called a proto-mission capable, pMC) spacecraft is summarized in Table~\ref{tab3a} and a notional drawing of the MC in one particular configuration is shown in Fig.~\ref{fig:sc-concept} with thruster locations shown but not in the configuration for full 3 axis control.  The MC comprises two pMC modules. The SGL spacecraft size will be determined by the optical telescope primary mirror which will be used for science data collection as well as for communication and navigation purposes. Following sail deployment, the sailcraft will use on-board propulsion or solar sails that can articulate to reduce the heliocentric      velocity and follow a flight path toward a close solar flyby.  This places the pMC vehicles on a solar system escape trajectory with an exit state vector toward the pre-determined SGL image position. If in-flight assembly is used, because of the difficulties in producing very large sails, the spacecraft modules (i.e., pMCs) are placed on separate sailcraft.  After in-flight assembly, the optical telescope and if necessary, the thermal radiators are deployed.  Analysis shows that if the vehicle carries a tiled RPS (green in Fig.~\ref{fig:sc-concept}) where the excess heat is used for maintaining spacecraft thermal balance, then there is no need for thermal radiators.  The MCs use electric propulsion (EP) to make all the necessary maneuvers for the cruise ($\sim$25 years) and science phase of the mission. The propulsion requirements for the science phase are a driver since the SGL spacecraft must follow a nonlinertial motion for the 10-year science mission phase.

\subsection{\label{sec:4.2}Measurement  Description}

The science objectives require that the telescope sample a projected image that is kilometers or tens of kilometers in size. Moreover, the projected image is always in constant movement as a result of the combined motions of the Sun (i.e., the lens), the exosolar system and the exoplanet itself.  The telescope must collect photons image pixel by image pixel.  As data from a growing number of image pixels are acquired, they can be used to reconstruct images of the exoplanet at increasing  spatial or spectral resolution.  Finally, to avoid motion blur that would make image reconstruction impossible, the imaging instrument must remain reliably ``on pixel'' for the duration of the pixel integration time. This requires a pixel-to-pixel relative navigational accuracy at meter-scale or better.

Regardless of the distance the images are being taken, there is noise.  1) The Einstein ring carrying the desired exoplanet photons will be disguised by leakage light from the solar corona. Even if the corona brightness is measured independently and accurately, the random shot noise due to corona photons needs to be compensated by   sufficiently long integration times. 2) The instrument sensors must be sensitive to very small variations in brightness in the faint Einstein ring seen on top of the bright corona background. For a 1-m class telescope, the corona surface brightness exceeds the surface brightness of the Einstein ring by a factor of $\sim3\times 10^4$ or greater \cite{7-Turyshev-Toth:2020-extend,10-Toth-Turyshev:2020}.  We estimate at least 16 bits of a 24-bit sensor will be solar corona light.  Finally, because these background noise sources need to be subtracted, there is a further requirement of the operational tolerances of the sensors: The coronagraph must be sufficiently uniform to enable calibrations by subtraction.

To deliver as much usable science as possible, the SGL instrument needs to obtain views of the Einstein ring that    are both spatially and spectrally resolved. Such data will make it possible to apply different image reconstruction techniques, including conducting trade-offs between spatial and spectral resolution, as well as progressively improving the image quality as more observational data become available. All this requires, that the raw observation data be returned to Earth. The minimum required data rate can be calculated based on the chosen integration time, the spatial resolution  at which time the Einstein ring is observed (constrained by the diffraction limit of the imaging instrument), the dynamic range of the    analog to digital converter of the imaging instrument and the number of spectral channels.

\subsection{\label{sec:4.3}System Description}

\subsubsection{\label{sec:4.3.1}Payload}

\begin{figure}[h!]
%  \vspace{-20pt}
\centering\includegraphics[width=0.60\textwidth]{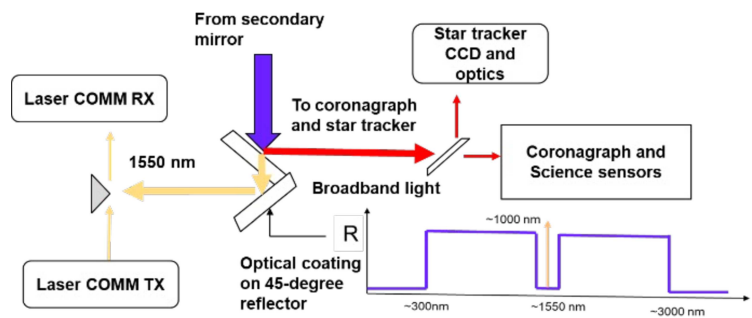}
  \caption{Block diagram of the SGL integrated payload instrument.}
\label{fig:instrum}
  \vspace{-10pt}
\end{figure}

The payload design merges several functions to reduce mass.  The primary payload comprises a coronagraph and sensor, the laser communications transmit (TX) and receiver (RX) units and a star tracker.  The light collector for all these subsystems is the telescope primary mirror.  To reduce mass the mirror utilizes replicated optics technology which is in current development. Prototype mirrors have been fabricated and tested in laboratory. A functional block  diagram of the integrated payload instrument is shown in Fig.~\ref{fig:instrum}.

%%%%%%%%%%%%%%%%%%%%%%%%%%%%%%
%%%%%%%%%%%%%%%%%%%%%%%%%%%%%%

The primary payload also implements a miniature diffraction-limited high-resolution spectrograph which further parses the already low photon flux.  A mitigation approach for increasing the photon flux is a larger telescope aperture, a coronagraph with blocking capability greater than $10^{-7}$ or longer integration times. At 1 $\mu$m, the light amplification of the SGL is $\sim2 \times 10^{11}$, so an exoplanet, which is initially seen as an object of 32.4 mag, becomes a $\sim$4.9 mag object. When averaged over a 1-m telescope, the light amplification is reduced to $\sim2\times 10^9$ \cite{4-Turyshev-Toth:2017}, resulting in a brightness of 9.2~mag, which is sufficiently bright (even on the solar corona background).

\subsubsection{On-board power}

The mission duration and the distances from which communications must occur primarily drives the requirements on the communication approach used.  The size of the power source is important but the RPS size is primarily driven by the need for propulsion to maintain the non-inertial trajectory. In this report we have analyzed the extreme case in which the propulsion is nearly always on, with the spacecraft continually following the exoplanet image. An   alternative approach would be flying inertial segments with intermittent propulsion.  Continuous electric propulsion draws multiple watts of power.

\begin{figure}[h!]
\centering \includegraphics[width=0.30\textwidth]{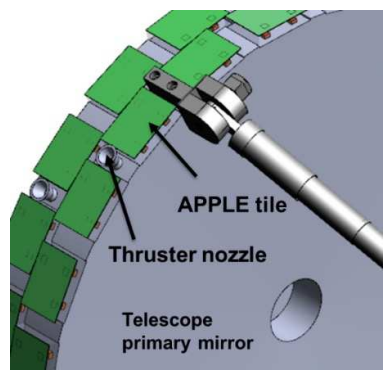}
  \caption{An SGL vehicle concept using the APPLE tiles (green) around the circumference for power energy storage and heating.}
\label{fig-apple}
  \vspace{-10pt}
\end{figure}

Table~\ref{tab4} shows the mass and average power budget results for two SGL configurations having different power sources and under two science phase scenarios.  It includes the calculated non-inertial trajectory path including position and acceleration to evaluate the power needs. The two science phase scenarios are to 800 and 900 AU with RPS that is either a \textonehalf-scale Brayton or RPS tile system called APPLE for ``Atomic planar power for lightweight exploration'' (APPLE) \cite{22-Nemanick-etal-2022}. The analysis includes a small systems reserve contingency for bus mass and bus power. The Brayton-type RPS is chosen because of the larger thermoelectric conversion efficiency ($\sim$20\%) that is available over conventional radioisotope thermoelectric generators (6-10\%). APPLE is a low TRL technology under development and integrates a radioisotope heat source, a thermoelectric generator and a radiation hard solid state battery. The integrated structure is a tile that can be mounted on the outside of spacecraft.  A notional design of an APPLE based spacecraft is shown in Figs.~\ref{fig:sc-concept} and \ref{fig-apple}. For a Brayton RPS spacecraft, the stack height is double.

%%%%%%%%%%%%%%%%%%%%%%%%%%%%%%
%%%%%%%%%%%%%%%%%%%%%%%%%%%%%%

The analysis results in Table~\ref{tab4} show a large variation in the space vehicle (SV) mass (power) varying from 42 kg (52 W) to 151 kg (151 W) depending on the RPS choice and mission length. These values are for an MC spacecraft that have the capability to transmit 10 Gbit/year at 1 kbps.

\begin{table}[h!]
\begin{center}	
\caption{\label{tab4} Mass and power budget results for two SGL SV configurations with a Brayton and APPLE RPS and under two science duration scenarios. Note: 1) SV mass and power after solar sail deployment, 2) Power budgets are averages  include operational modalities.}
{\small
\begin{tabular}{|p{0.37\textwidth}|>{\centering}p{0.09\textwidth}|>{\centering}p{0.09\textwidth}|>{\centering}p{0.10\textwidth}|>{\centering}p{0.10\textwidth}|}
\hline
& To 800AU APPLE-RPS
& To 900 AU APPLE RPS
&To 800 AU \textonehalf-scale Brayton RPS
&To 900 AU \textonehalf-scale Brayton RPS
\tabularnewline
\hline
 & SV & SV & SV & SV
 \tabularnewline
\hline
{\bf Bus Mass (kg)} & {\bf 33.9} & {\bf 41.3} & {\bf 86.2} & {\bf 140.6}
\tabularnewline
Attitude Determination and Control (kg) & 2.7 & 2.7 & 5.4 & 5.4
\tabularnewline
Command and Data Handling (kg)
&0.4 & 0.4 & 0.8 & 0.8  \tabularnewline
Communications/TT\&C (kg)
&4.7 & 4.7 & 8.8 & 8.8
 \tabularnewline
Electrical Power (kg)
&8.8 & 10.1 & 31.4 & 40.4
 \tabularnewline
Harness (kg)
&0.7 & 0.8 & 2.8 & 3.8
 \tabularnewline
Propulsion (kg)
& 6.6 & 11.0 & 13.2 & 46.4
 \tabularnewline
Structure/Mechanisms (kg)
&2.7 & 2.7 & 5.3 & 5.3
 \tabularnewline
Thermal Control (kg)
& 0.5 & 0.6 & 1.2 & 1.5
 \tabularnewline
Systems Reserve Contingency (kg)
&6.8 & 8.3 & 17.2 & 28.1
 \tabularnewline
{\bf  Payload Mass (kg)}
& {\bf 7.7} & {\bf 7.7} & {\bf 10.3} & {\bf 10.3}
\tabularnewline
Solar Sail/Radiator
& 1.0 & 1.0 & 1.0 & 1.0
\tabularnewline
Robotic Arm + Secondary Mirror
& 1.6 & 1.6 & 1.6 & 1.6
\tabularnewline
Primary Mirror
& 2.0 & 2.0 & 2.0 & 2.0
\tabularnewline
Docking Mechanism + Power \& Data Interface
& 2.0 & 2.0 & 4.0 & 4.0
\tabularnewline
Coronagraph
& 0.6 & 0.6 & 0.6 & 0.6
\tabularnewline
RPO Sensor
& 0.6 & 0.6 & 1.1 & 1.1
\tabularnewline
{\bf SV Mass (kg)}
& {\bf 41.6} & {\bf 49.0} & {\bf 96.5} & {\bf 150.9}
\tabularnewline
{\bf Bus Power (W)}
& {\bf 51.5} & {\bf 60.2} & {\bf 98.1} & {\bf 151.2}
\tabularnewline
Attitude Determination and Control (W)
& 3.3 & 3.3 & 3.3 & 3.3
\tabularnewline
Command and Data Handling (W)
& 2.1 & 2.1 & 3.1 & 3.1
\tabularnewline
Communications/TT\&C (W)	
& 13.6 & 13.6 & 13.6 & 13.6
\tabularnewline
Electrical Power (W)
& 1.5 & 1.8 & 3.0 & 4.6
\tabularnewline
Propulsion (W)
& 20.3 & 26.9 & 54.5 & 95.8
\tabularnewline
Structure/Mechanisms (W)
& 0.0 & 0.0 & 0.0 & 0.0
\tabularnewline
Thermal Control (W)
& 0.5 & 0.5 & 1.1 & 0.5
\tabularnewline
System Reserve Contingency (W)
& 10.3 & 12.0 & 19.6 & 30.2
\tabularnewline
{\bf Payload Power (W)}
& {\bf 5.8} & {\bf 6.3} & {\bf 8.6} & {\bf 9.9}
\tabularnewline
Solar Sail/Radiator
& 0.0 & 0.0 & 0.0 & 0.0
\tabularnewline
Robotic Arm + Secondary Mirror
& 0.0 & 0.0 & 0.0 & 0.0
\tabularnewline
Primary Mirror
& 0.0 & 0.0 & 0.0 & 0.0
\tabularnewline
Docking Mechanism + Power \& Data Interface
& 2.8 & 3.2 & 5.5 & 6.9
\tabularnewline
Coronagraph
& 3.1 & 3.1 & 3.1 & 3.1
\tabularnewline
RPO Sensor
& 0.0 & 0.0 & 0.0 & 0.0
\tabularnewline
\hline
Power Generated (BOL), W
& 110.4 & 127.2 & 220.0 & 275.0
\tabularnewline
Power Generated (EOL), W
& 80.5 & 89.1 & 160.4 & 192.7
\tabularnewline
\hline
\end{tabular}
}
\end{center}
\vskip -10pt
\end{table}

\subsubsection{\label{sec:4.3.2}Flight System}

The SGL spacecraft includes all the necessary systems for delivering and supporting the optical telescope to the SGL's focal region. The spacecraft structure is a simple cylinder-like construction to accommodate the payload, spacecraft electronics, propulsion, attitude control, communications, thermal, and power subsystems. The estimated mass by subsystem is given by Table~\ref{tab4}. The trade study that produced the results in Table~\ref{tab4} included several operating modes, e.g., data acquisition, communications to Earth and propulsion.  The various operating modes were implemented to explore means by which to reduce the total power requirements.  For the science phase, propulsion was continuously ON, further refinements could be explored in a scenario where there is thrust then coast.  In the cruise phase, attitude control is done by field emission electric propulsion (FEEP) thrusters which minimize the cost/mass of propellant tanks since there is no stringent pointing requirement for the spacecraft.

\begin{figure}[hbt!]
\centering
\includegraphics[width=.95\textwidth]{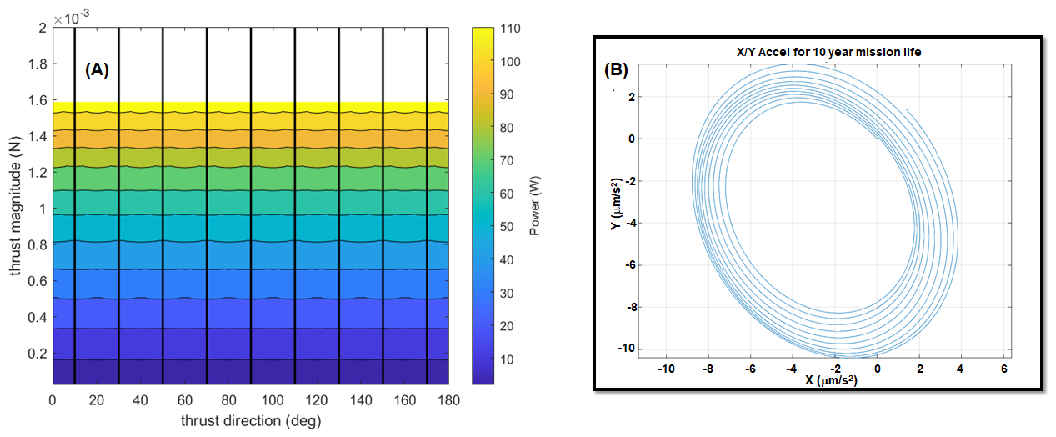}
\includegraphics[width=.45\textwidth]{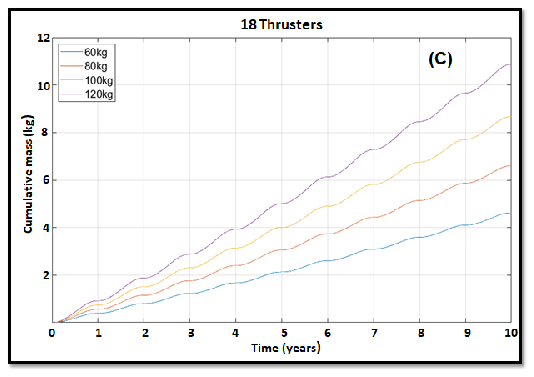}
\caption{(A) A 2D thrust control simulation using a uniform distribution of 18 FEEP thrusters on a 120 kg notional vehicle.   (B) The cross-track accelerations from the science phase trajectory calculations used in the simulation. (C) Cumulative fuel mass as a function of science phase duration for various notional vehicle masses.
}
\label{fig5}
\end{figure}

The attitude control sensors include standard star trackers, inertial measurement units, coarse sun sensors and use of optical astrometry.  In the science phase the attitude control sensor uses the shape of SGL image on the sensor while FEEP thrusters with colloid propellants are used to make precise trajectory changes. Fig.~\ref{fig5}(a) shows the results of thrust control simulation given a uniform distribution of 18 thrusters about a notional vehicle of 60-120 kg mass.  Fig.~\ref{fig5}(b) is the acceleration input to the simulator (the critical motion path) in the cross-track direction.  The results show that as a function of thruster direction the magnitude is uniform over a wide range of thruster input power.   Finally, in Fig.~\ref{fig5}c we plot the cumulative fuel mass as a function of science phase duration for various notional vehicle masses.

The baseline design for the command and data handling subsystem is based on RAD750 processors that interfaces to the attitude control, telecommunication, power, and payload subsystems, with sufficient memory storage  to accumulate up to two weeks of instrument data.  However, this investigation has not concluded the amount of onboard autonomy and computational processing that will be needed.  The RAD750 may not be sufficient and high-performance processors maybe necessary. Given that these high-performance units run at high speed ($>$GHz), they are likely to be more susceptible to radiation effects. Similarly  anticipated for the onboard memory.  Layered adaptable systems need to be developed at various levels of the electronic systems packaging hierarchy possibly down to the active transistor level.

\subsubsection{Communications}

As currently conceptualized, the primary communications link to Earth is via optical means. Laser communications (1550 nm, 60 kHz linewidth) is the clear choice for   communication from 900 AU. Laser communications are now being deployed in space systems   for both crosslink and downlink applications\footnote{NASA just completed a 10-year technology demonstration project (Laser Communications Relay and Demonstration: LCRD) lead by NASA/Goddard with a goal to move TRL from 6 to 8.} \cite{24-Rose-etal:2018}. Our communications link budget analysis shows that with a 15W optical power laser, 1-m Tx, and Rx telescopes, it is possible to transmit 9bps (4dB margin, 116\_PPM encoding) from 900 AU, which defines the end of the SGL mission. This analysis  is based on a small constellation of near Earth, in-space, 1-m aperture Rx telescopes and should only be implemented if all the information can be encapsulated in the return of a few bits (i.e., all computation done onboard).  At the upper end, the science requirements estimate a data capture rate of 16 Gbits/year assuming an exoplanet-image availability of 70\% per year.  This resulting image comprises $\sim 5.5$~K pixels which is parsed into 100 spectral channels.  If all the data bits are returned to Earth, then for a single spacecraft and a 1~kbit/s communication rate (no image compression) a laser (ON/OFF) duty cycle of $\sim50$\% is necessary.   At 8~kbit/s communication rate the duty cycle reduces to 6.4\%.   Given the baseline Tx optical communication system (15~W laser, 1-m Tx telescope), higher throughput can be achieved by using larger Rx apertures.

\begin{table}[h!]
\begin{center}
\caption{\label{tabV} Results of a laser communication link from a single spacecraft from 900 AU using a 15 W laser.}
\begin{tabular}{|>{\centering}p{0.08\textwidth}|>{\centering}>{\centering}p{0.08\textwidth}|>{\centering}p{0.08\textwidth}|>{\centering}p{0.08\textwidth}|>{\centering}p{0.10\textwidth}|>{\centering}p{0.08\textwidth}|>{\centering}p{0.12\textwidth}|p{0.08\textwidth}|}\hline
Laser Power, [W] & Range, [AU] & Bit Rate, [bps]
& Link margin (dB) & Tx telescope aperture (cm) &Rx telescope aperture (cm)& Rx Pointing (nanoradians)	& Telescope efficiency\\\hline\hline
15 & 900 & 200 & 4.1 & 100 & 500 & 50 & 50      \\
15 & 900 & 1000 & 4 & 100 & 1000 & 10 & 50      \\
15 & 900 & 8000 & 4.4 & 100 & 3000 & 50 & 50      \\
\hline
\end{tabular}
\end{center}
\vskip -10pt
\end{table}

Table~\ref{tabV} shows the link budget for ground station receivers with different apertures.  The results show 1 kbps transfer rate from 900 AU is possible given a 10~m class ground telescope. An alternative scenario is for multiple spacecraft transmitting separate segments of the data back (e.g., 5 spacecraft each communicating at 200 bps could be supported by a 5~m diameter ground telescope).  If lossless data compression techniques are applied, for example a factor of 2, then a single spacecraft operating at 1kbps or 5 operating at 200 bps could in each case, deliver all the data to the ground during 30\% of the year (3.6 months) that the exoplanet is not visible.  The link margin analysis includes atmospheric losses  (wavefront error/Strehl, scintillation, absorption, Mie-scattering, high Cirrus clouds) which totals over 5 dB. In addition, the calculation includes background noise (e.g., stellar), detector dark and thermal noise, and optical amplifier (LNA) noise.   Table~\ref{tab6} presents further details of the analysis for the case of 30 m ground telescope.

\begin{table}[h!]
\begin{center}
\caption{\label{tab6} Details of the analysis for the case of the 30~m ground telescope; communication rate is 8~kbps.
}
\centering
\includegraphics[width=.97\textwidth]{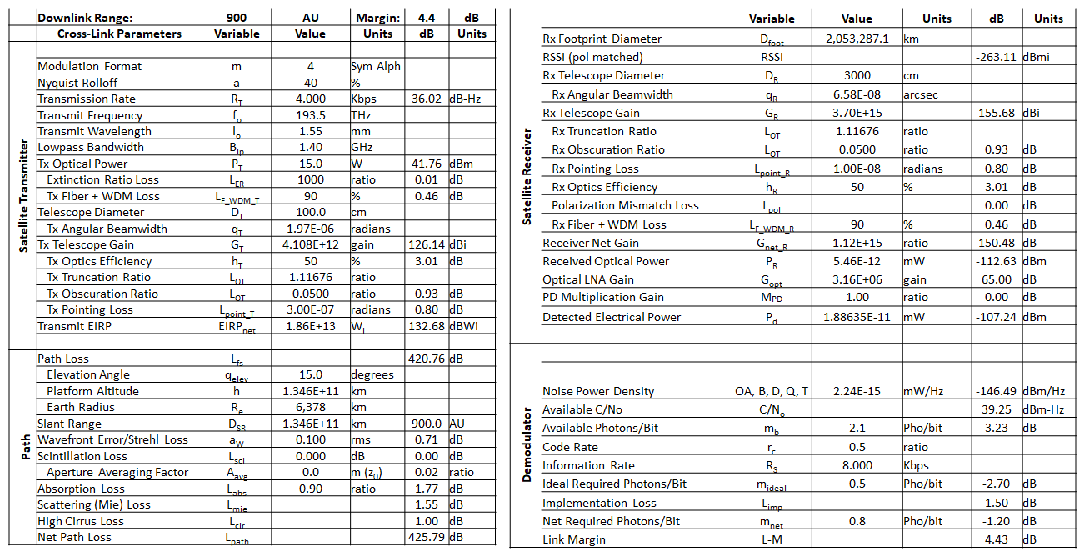}
\end{center}
\vskip -10pt
\end{table}

While optical COMM is the method for communications to Earth, among the cluster of the SGL spacecraft, RF communication is more viable, and a local area network is envisioned in the design. With transmit powers on the order of 1 W, a data rate of 10 kbps for Ka band (2.8 GHz) is possible (up to $5\times 10^4$ km separation) or 1 Mbps at $2\times 10^4$~km separation for V band (80 GHz). At closer ranges and V-band, 10--100 Mbps is possible at ranges of $(2-6)\times 10^3$~km (3dB margin). These calculations are important because they help to define the kind of distributed processing and control that might be implemented for a cluster of free-flying SGL spacecraft. This is also critical in the science phase where a formation flying cluster of SGL spacecraft must maintain precise distances while executing a complex trajectory.

\subsubsection{Thermal management}

The thermal requirements for the SGL mission are unique. The spacecraft must endure extreme solar heating while  operating in the vicinity of the Sun. After perihelion, its solar sail may double as a heat shield for the electronics. Later, the spacecraft must survive in the deep space environment.

The unique aspect is that the mission broaches two temperature extremes ($-190{}^\circ$C to 250${}^\circ$C) and its effect on the materials used must be studied, but more so is the effect on electronics.  Work done at NASA/Glenn shows that the switching characteristics of a MOSFET device or a DC/DC converter varies with temperature \cite{[26]}. Consequently, circuit designs must be implemented that permit recalibration of components.  A possible thermal management solution is to run the spacecraft hotter than standard, which could reduce periodic recalibration of electronics.  In addition, it would also increase the off-gas rate from materials and thereby reduce the overall ambient contamination level.   The most significant heat sources in the space vehicle are the RPS, the EP thrusters, when operating, and the laser transmitter.  For the case of a Brayton RPS, which is a localized heat source and continuously ON ($\sim$300~W thermal for 50~W electric), the thermal management is by heat pipes to external radiators.  An attempt to distribute the heat from a large central source adds mass.  In the case of a distributed RPS, like the APPLE-tile ($\sim$20~W thermal for 2 W electric) the local material could serve as insulation.   Finally, the overall structure shape of the SGL vehicle is a round cylinder (Fig.~\ref{fig:sc-concept}) and while not shown, the intent is to make it as an open scaffold.  Thermal management is then heat transfer through the scaffold rods and radiation.

\subsubsection{\label{sec:4.3.3}Ground Segment  and Mission Operations Systems}

One or more large Earth observatories will be assigned for communications with the SGL spacecraft. The ground stations will use lasers very similar to those on SGL vehicles.  The ground station lasers should be operating at 1550nm while the space based system could operate at the more mature technology 1064 nm laser wavelength.  The difference arises because the space based sensor is essentially staring at the Sun (albeit behind a coronagraph) and at 1550 nm the signal to noise against the solar light is better than at 1064nm. Of more importance is the uplink linewidth of the laser.  The ratio of SNR(1550)/SNR(1064 nm), where SNR is against solar radiation, show values of 1.4, 1.3 and 1.2 for laser linewidths (FWHM) of 33, 60 and 400 kHz  respectively. Consequently, it would be beneficial if the uplink high power laser had a narrow emission linewidth with the need of a narrow linewidth filter on the space vehicle Rx optics as well.

The onboard clocks will need to be periodically synchronized with Earth clocks. At 20 AU/yr velocity, time dilation is $\sim$1.6 sec/yr.  The SGL design maintains the on-board clocks synchronized to Earth by $<1$ ppm for the duration of the mission.   The baseline clock is the chip scale atomic clock (e.g., Microsemi, $<$120 mW, which is rad tolerant to 20 krad and SEU-64 MeV cm${}^2$/mg and LET tested for cosmic rays -- 10--100 MeV cm${}^2$/mg).  It is at TRL 7-8, but at the current Allan deviation values, it requires synchronization every 4 months for a 10 MHz clock.  In the future, we envision the use of optical frequency comb clocks (currently TRL 2-4) and given the current Allan deviation values, no adjustment would be necessary for 1300 years for clocks running at 10 MHz or 32 years at 500 MHz.  We anticipate a 10X improvement in the optical frequency comb clocks within a decade which will permit synchronization-free operation for 53 years for a 3GHz clock.  Given the roughly 1-week round trip communication time, the science operations will have to be autonomous with minimal control from the ground. The amount of autonomy for solving system anomalies still remains needs to be studied.  The spacecraft could have two view   periods each week. Observing times are unconstrained. Times will be selected a week in advance and once a week the spacecraft will communicate with the optical Deep Space Network (DSN) to download the previous week of science data and to upload the schedule of tracking times for the following week.

\subsection{CONOPS for launch and cruise phases}

There are three primary phases of the mission, Launch (liftoff, deployment, dive toward perihelion, in-space assembly), Cruise and Science.  The CONOPS for the launch and cruise phases of  the mission, while technically challenging themselves, do not alter approaches already in use. A solar sail propulsion  system will have to be developed that can sail to a particular perihelion location. The design of the attitude control system (ACS) of the solar sail propulsion is one challenge to overcome given the close interrelationship between thrust vector and attitude toward the sun. To allow for effective in-space assembly, the sailcraft must fly in formation to arrive together at the rendezvous point.

Position and velocity requirements for the incoming trajectory prior to perihelion are $< 1$~km and $\sim$1~cm/sec \cite{Garber:2021}. Timing through perihelion passage is days to weeks with errors in entry-time compensated in the egress phase. As an example, if there is a large position and/or velocity error upon perihelion passage that translated to an angular offset of 100'' from the nominal trajectory, there is time to correct this translational offset with the solar sail during the egress phase all the way out to the orbit of Jupiter. The sails lateral acceleration is capable of maneuvering the sailcraft back to the desired nominal state on the order of days depending on distance from the Sun. This maneuvering capability relaxes   the perihelion targeting constraints and is well within current orbit determination knowledge threshold for the inner  solar system which drive the $\sim$1 km and $\sim$1 cm/sec requirements.

The CONOPS for the in-space assembly phase are known.  There have been multiple demonstrations of autonomous in-space docking to form  larger entities. Proximity operation technologies (i.e., sensors, fiducials, approach algorithms) exist and will be further refined with time\footnote{NASA's space technology mission directorate has already started such a project (On-Orbit Autonomous Assembly from Nanosatellites-OAAN) and has followed with a CubeSat Proximity Operations Demonstration (CPOD) mission. }  \cite{[28]}.

The CONOPS of the cruise phase entails the use of the primary 1-m telescope for navigation based on optical astrometry. The telescope looks back at the solar system planets. In track navigation is via direct communication link. Cross track is via optical parallax against a background star field. Our analysis shows that accuracy of $< 10^4$~km can be met at 550 AU. This is a conservative analysis (i.e., no a priori state information) and uses combined data from optical navigation and parallax processing.   A 30-day fit span is used with 1- and 5-minute collection times for 0.01'' angular resolution ($1\sigma$). Table~\ref{tabVII} shows the  state covariance results using a combination of the stars and gas giants.

\begin{table}[h!]
\caption{\label{tabVII}Estimated state covariances at 550 AU ($1\sigma$) \cite{Garber:2021}.
}
\vskip -12pt 
\centering\begin{tabular}{|l|>{\centering}p{0.09\textwidth}|>{\centering}p{0.1\textwidth}|>{\centering}p{0.08\textwidth}|>{\centering}p{0.08\textwidth}c|}
\multicolumn{6}{c}{~}
\\\hline
\multicolumn{6}{|c|}{Data taken at 5-minute rate (0.01'' angular resolution)}\\\hline\hline
~&RSS (km)&Range (km)&RA (km)&dec (km)&\\\hline
Jupiter and Saturn&9403.5&9403.1&68.1&55.1&\\
Jupiter, Saturn, stars&8905.1&8905.0&14.3&24.7&\\
Jupiter, Saturn, Neptune&2792.9&2792.1&54.8&43.7&\\
All combined&2611.2&2610.8&34.9&31.8&\\\hline
\multicolumn{6}{|c|}{Data taken at 1-minute rate (0.01'' angular resolution)}\\\hline\hline
Jupiter and Saturn&4278.8&4278.6&30.4&24.5&\\
Jupiter, Saturn, stars&3983.1&3983.0&6.42&11.1&\\
Jupiter, Saturn, Neptune&1249.11&1248.8&24.5&19.5&\\
All combined&1170.7&1170.5&15.6&14.0&\\\hline
\end{tabular}
\end{table}

\subsection{CONOPS during science phase}

The CONOPS for the science phase is a bit more challenging. A meter-class telescope cannot ``see'' the resolved exoplanet. The projected image of the exoplanet, as it would appear on an imaginary movie screen is several square kilometers in size. Instead, the telescope looks at the Sun, observing the Einstein ring around the Sun and measures changes in its brightness as the telescope traverses the image plane.

The actual data that results in a single image pixel consists of a number of captured exoplanet photons, not an image. A picture of the exoplanet only arises when multiple, time-and-position stamped image pixels are obtained and via deconvolution methods reconstruct the image of the source. To acquire photons that can be sourced to the exoplanet, the mission requires very precise knowledge of position: Image-pixel to image pixel positioning accuracy must be on the order of 1-m  \cite{7-Turyshev-Toth:2020-extend} or better.

The image pixel is also in motion due to the solar wobble (acceleration of $6 \times 10^{-6}~{\rm  m/s}^2$)) which is the largest effect.  The reflex motion of the host star is $< 3\times 10^{-11}~{\rm m}/{\rm s}^2$ per year.  Consequently, the SGL spacecraft must make periodic adjustments in the trajectory as they fly out from 650 AU.  The multi-year trajectory path of the SGL spacecraft resembles a spiral of increasing radius when viewed in 3D space \cite{14-Turyshev-Toth:2022-wobbles}.

The acceleration requirements are not difficult to achieve given the onboard EP thrusters, it is the metrology   during the motion that is challenging. The CONOPS for searching and finding the POA of the host star use the change in the image shape of the host star on the SGL spacecraft sensor as a navigation tool. The change in the host  star image as the SGL spacecraft approaches the POA of the host star has been modeled and reveals a sequential set of definable images which can be used as guide  \cite{7-Turyshev-Toth:2020-extend}.   The general conclusions from the model are summarized below.
{}
\begin{itemize}
\item Perhaps as a comforting beacon, the host star will always be visible on the payload telescope during the cruise phase.  It appears as an unmagnified star. In the science phase, the CONOPS entails moving the spacecraft in the cross-track direction to bring the unmagnified  image closer to the center of the coronagraph-stop.
\item While maintaining the cross track motion, the CONOPS further entails bringing the amplified spot image, which develops with proximity,  even closer to the  coronagraph-stop edge.
\item As the amplified spot image approaches the coronagraph-stop edge, a second amplified image appears 180$^\circ$ opposite the first. Both spot images straddle the coronagraph-stop edge.  There is no reason to move on to the POA of the host star, it is only necessary to know its location.
\item If there were no optical background noise, the change in shape of the light on the CCD as the spacecraft approach the POA has also been simulated. First, the intensity of both spots grow. Then, the two spots change shape to form small arclets around the coronagraph     edge, the arcs grow to form a perfect ring about the coronagraph edge when on the POA.
\end{itemize}

\begin{figure}[t!]
\centering
\includegraphics[width=0.40\textwidth]{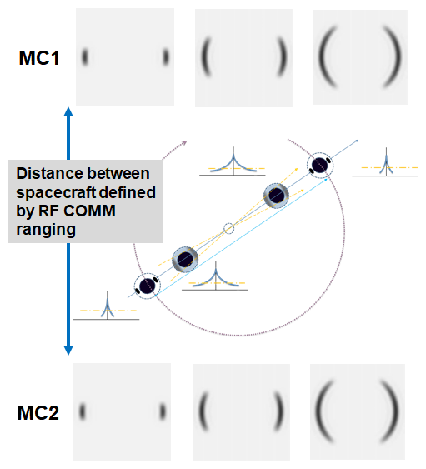}
\hskip 20pt
\includegraphics[width=0.552\textwidth]{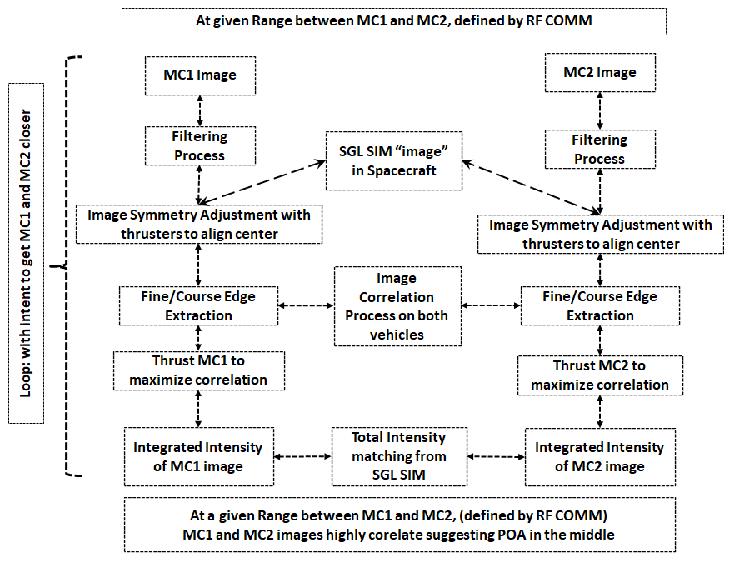}
\caption{\label{fig6} Left: Schematic of the CONOPS used to find the POA of the host star using two spacecraft.  Right: Flowchart of the image correlation analysis used to ensure centering on the host star POA.}
\end{figure}

In the case of searching the POA of the host star, where SNR is high, the change in the image shape permits a CONOPS to be devised that use image processing and cross-track angular motion to advantage. Figure~\ref{fig6} presents a schematic of the CONOPS. On the left, 3 images are shown that would be observable on two SGL spacecraft (i.e., MC1 and MC2) as each approaches the POA of the host star from 180 degrees opposite. The images are from the simulation  \cite{10-Toth-Turyshev:2020,7-Turyshev-Toth:2020-extend,14-Turyshev-Toth:2022-wobbles} as described above. Also shown in the middle of left panel are representations of the sensors that are on the two SGL spacecraft facing each other while in a slow cross-track rotation maneuver. Range distance between MC1 and MC2 is measured via relative RF navigation. Each spacecraft exchanges image data with its counterpart. Image correlation processing is done by both spacecraft (in the flow chart shown on the right panel) and the data is used to fire thrusters.  The right panel schematically describes the process steps from image capture, filtering, symmetry alignment and edge extraction to correlation analysis resulting in the command to fire key thrusters that make the images match (i.e., increase image correlation from sensors in MC1 and MC2).  The image processing and the synchronized rotational motion of  MC1 and MC2 enables the centering of the POA between the two spacecraft.  Only two SGL spacecraft are necessary to find the POA of the host star. The localization of the host star POA could take as long as 5 years (i.e., from 550 AU to 650 AU if traveling at 20 AU/yr).

\subsection{CONOPS of the exoplanet search}	

The CONOPS for finding the POA of the exoplanet follows a similar approach as that for the host star, except that the estimated light intensity from the host star is $10^{4}-10^{5}$ larger than that from the exoplanet.   Moreover, some of the amplified host star light is always present and appears as two spots, as the SGL spacecraft journey toward the exoplanet.  Finally, leakage of solar light from coronagraph appears as background ``noise'' on top of the light from exoplanet. Consequently, the CONOPS for finding and ``locking onto'' the exoplanet image is a bit more complex.  The key points are described below.
\begin{itemize}
\item The intensity of the solar corona can vary along with its size as the spacecraft travel from 650--900 AU ($\sim$10 years).  For example, a maximum extension of up to $12R_\odot$ has been recorded. Consequently, the solar corona light has to be subtracted during the data recording phase.
\item Our SGL architecture currently comprises 5 SGL spacecraft all capable of conducting the mission. To remove the effects of the solar corona, one SGL spacecraft is directed on an inertial path down the center of the spiral trajectory while the 4 remaining SGL spacecraft move in the noninertial frame. Having a spacecraft moving in an inertial frame serves two purposes: i) Because the spacecraft is far from the POA  of exoplanet, the light on its sensor is only the solar corona leakage light from the coronagraph and the light leakage from the host star.  No exoplanet photons should be present. Consequently, the image from the inertial travelling frame spacecraft can be subtracted by the 4 in the noninertial frame. ii) the SGL spacecraft traveling the inertial path serves as a local reference frame and can be used for navigation.
\item The analog to digital converter on the spacecraft sensor should have large dynamic range ($>$24 bits), because at least 16 bits of the dynamic range will be solar corona light.  This knowledge places a requirement that the relative accuracy of the coronagraphs should be $\ll$ 0.4\% to make background subtraction possible.  This could be achieved by holding to high manufacturing standards and to periodic in space calibrations among the 5 SGL spacecraft.
\item There are two approaches to remove the amplified ``leakage'' host star light from the image sensors.  1) Electronically subtract the pixels that are saturated with the leakage light and 2) use a digital mirror device (DMD) in the payload sensor optical train to physically keep the leakage light from hitting the sensor pixels.  The former is simpler but susceptible to pixel burnout because one wants the gain to be high on the pixels not sensing the host star light.  In the latter case, it requires insertion of a MEMS chip that comprises a Mpixel count of micromirrors, which can be individually turned OFF for the sensor pixels that have large intensity.  A DMD can also be used to increase the dynamic range of a CCD camera by 48 dB (8 bits) \cite{36-Feng-etal:2017}. Multiple DMDs can be ganged to further increase the dynamic range. The increase is related to how long a DMD pixel-mirror remains ON. Current DMD mirrors can modulate at 32 kHz (in binary mode). A DMD has also been used for wavefront modulation maximizes quantization or spatial resolution  \cite{37-Georgieva-2020}. Given the beneficial features of the DMD, it does however increase the complexity of the optical design.
\end{itemize}

\begin{figure}[t!]
\centering
\includegraphics[width=.70\textwidth]{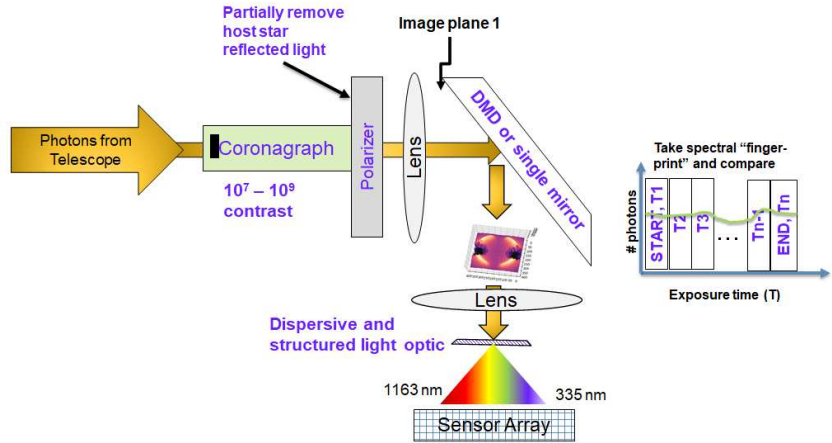}
\caption{Description of the optical train for data acquisition and analysis.
}
\label{fig7}
\end{figure}

Figure~\ref{fig7} shows a schematic of the optical train for data acquisition where light from the telescope passes through the coronagraph, polarizer and is imaged on the DMD or a simple mirror surface.

The CONOPS of the data acquisition requires that the relative positions of the data acquiring spacecraft be known with very high accuracy.  This is accomplished by use of the RF communications (e.g.,  V-band, $\lambda \simeq$~4~mm) as ranging or by a laser modulated cross-link scheme (e.g., phase shift optical carrier phase modulation). The latter is a coherent heterodyne technique that provides both range and velocity information with accuracies of 1 mm and 4~mm/sec respectively \cite{34-Le-etal:2020}.  The position-placement of the data-acquiring spacecraft is referenced to the lone spacecraft flying the inertial path which also keeps track of the image plane location.

Knowledge of the precise position of the spacecraft does not ensure that the collected photons during a sensor integration period comes from a single image pixel (called a frozen-frame data set). The location where each image pixel  was taken, must be known in relation to the moving 1300 m image plane which represents the whole exoplanet.  We anticipate that years of a priori ground telescope investigations, will refine the exoplanet orbit parameters to generate an accurate motion profile of the image plane, but to what accuracy?

During a single exposure on the sensor (e.g., 100 sec), the objective is to ensure that the information content on the image pixel remains constant (freeze-frame). We assume there is prior knowledge of the exoplanet rotation and its blurring effects on the image pixel can be removed by limiting the exposure time. For example, given an Earth size exoplanet 100 ly distant and a 150 km wide source image, the maximum exposure time to limit blurring due to rotation is 275 sec.  A CONOPS approach for ensuring a freeze-frame data set is to parse the time segmented captured light into spectral components.  On board sensor analysis tracks the spectral components that vary predictably with exoplanet rotation rate. The segmented-time data is then analyzed using a time-reversal algorithm to remove planetary rotation effects. If there are spectral components that vary differently in the segmented-time data, we surmise that the image frame is   not frozen and identify that image pixel as null.  The utility of spectral analysis depends on the estimated photon flux and shot noise, but it can be a very powerful tool.   Figure~\ref{fig7} (lower right) depicts in schematic form a time-parsed acquisition mode with spectral ``fingerprint'' comparisons.

Given sufficient photon flux, spectral analysis can also be applied to distinguish between a failure of an SGL spacecraft to hold a constant attitude during exposure with the overall drift of the exoplanet total image (i.e., trajectory path).  In the former case only the failed SGL spacecraft will yield a null image pixel while for the latter all 4 SGL spacecraft operating in the image plane will yield a null image pixel.

Time dependent events on the exoplanet shorter than the exposure time will yield a false negative. One mitigation approach is to parse the exposure time into shorter and shorter time segments; where a particular event measured over the course of one-time segment is marked and compared with that in the following timed segments to distinguish a null or frozen-frame.  These approaches strongly depend on sufficient flux of data-photons.

\subsection{\label{sec:4.4}Science Data Collection, Analysis, and Archive}

The primary data product from the SGL spacecraft will be a series of brightness data and positions where these measurements were taken. To the extent permitted by the communications capabilities, full-raw data (including background events) will sometimes be included so that it will be possible to verify the on-board compression efficacy. Auxiliary data such as on-board conditions, system health parameters, pointing information, etc. will make up the total dataset, but will constitute a small fraction of the data volume.

Data processing occurs in two stages. First, the  data are combined into meaningful imaging data, comprised of a representative brightness measurements, timestamps and spacecraft position. During this process, the on-board detector is calibrated, and various detector and timing system performances are verified. Second, having a set of imaging data from a variety of locations within the image, the measurements are compared to a highly developed, parameterized physical model of the exosolar system that includes gravitational dynamics and planetary physics influences. The difference between the model and actual data is minimized with respect to the parameters in the model; many of which are simply initial conditions for exosolar system bodies. A model for exoplanetary system is built to allow  exploration of departures from general relativity, which becomes the input to the spacecraft navigation with regards the SGL mission.

\subsection{Modularization and in-flight assembly}
\label{sec:mod-aasem}

A mission architecture has been investigated in which the mission payload and supporting instruments are parsed into modules where each is propelled to high velocities via solar sailcraft. Hence the need for an in-space assembly of distributed spacecraft segments to form a mission capable spacecraft.

\subsubsection{Modularization and in-flight assembly}

The modularized approach provides some advantages. 1)  The ability to utilize rideshare services as secondary payloads.  2) The use of containerized vehicles that have conventional propulsion.   The latter removes the time constraint as when to start the mission because the sailcraft can remain in a near Earth parking orbit protected until needed.  There is a trade-off on how far the containerized rideshare vehicles should carry the sailcraft toward perihelion versus  use of sailcraft propulsion to gather all SGL sailcraft into a viable formation proceeding to perihelion.

When sufficient containerized vehicles have been launched, the containers release the sailcraft which deploy solar sails and initiate a trajectory toward a pre-established solar perihelion point. Upon perihelion egress, the propulsive impact of solar sails diminish with distance. The solar radiation pressure at Earth orbit is less than 1\% of that at 15$R_\odot$ ($\sim 9~\mu$Pa vs 1.9 mPa).  We eject the large sails making it easier for the remaining 10--20 kg mass spacecraft to conduct docking maneuvers. This in-space assembly necessitates that each 10--20 kg mass unit be a fully functional nanosatellite, the pMC, and using its onboard EP thrusters docks to form a MC spacecraft.

\subsubsection{Nanosatellite bus}

The parsing of a MC into feasible units requires trade studies between function, size, weight and power. The trade space also needs to account for the mission length, the need for reliability and redundancy given the long mission, and  the need for near autonomous adaptability (e.g., machine learning) given the limited communications link to Earth.

To address these challenges, we relied on The Aerospace Corporation's Concept Design Center (CDC), which uses a comprehensive simulation approach based on the concurrent engineering methodology (CEM). An existing CEM tool \cite{McInnes:2021} was modified to explore the utility of fractionated spacecraft and includes features that prevision technology advances \cite{Stevens:2021}. Two types of distributed functionality were explored. A fractionated spacecraft system that operates as an ``organism'' of free-flying units that distribute function (i.e., virtual vehicle) or a configuration that requires reassembly of the apportioned masses. Given that the science phase is the strong driver for power and propellant mass, the trade study also explored both a 7.5 year (to $\sim$800 AU) and 12.5 year (to $\sim$900 AU) science phase using a 20~AU/yr exit velocity as the baseline. The distributed functionality approach that produced the lowest functional mass unit is a cluster of free-flying nanosatellites (i.e., pMC) each propelled by a solar sail but then assembled to form a MC spacecraft.

\begin{table}[h!]
\begin{center}
\caption{\label{tab3} Results of the trade study where a mission capable spacecraft is parsed into pMC modules using a \textonehalf-scale Brayton-type RPS unit}
\begin{tabular}{|>{\centering}p{0.20\textwidth}|
>{\centering}p{0.19\textwidth}|
>{\centering}p{0.16\textwidth}|
>{\centering}p{0.16\textwidth}|
p{0.14\textwidth}|}
\hline
4 pMC nanosatellites to form 1 MC &
\begin{flushleft}\vskip -1em
pMC mass = sailcraft mass 5 kg + $25\%$ mass contingency)
\vskip -2em \end{flushleft}
&
 \multicolumn{2}{c|}{Sailcraft area (m${}^2$) and on side (m)} &
Mass of an MC (with 25\%  mass contingency)\\\cline{3-4}
&&perihelion at 15$R_\odot$ &perihelion at 10$R_\odot$ &
\\\hline\hline
Science phase duration, ~7.5 yrs (800 AU)  & 29 kg &  4352 m${}^2$,  66 m  &   2610 m${}^2$,  51 m &
97 kg   \\
Science phase duration, 12.5 yrs (900 AU)   &  43 kg & 6453 m${}^2$,  80 m     &  3870 m${}^2$,  62 m &
151  kg \\
\hline
\end{tabular}
\end{center}
\vskip -10pt
\end{table}

Table~\ref{tab3} shows the results for the average mass of a pMC vehicle, the approximate sail area necessary to acquire   a 20 AU/yr velocity at two perihelion distances and the net mass for the mission capable spacecraft (in this case 4 pMC units because of type of Brayton RPS used). A ~25\% mass reserve has been included in the analysis. It is anticipated that technology advances could reduce these mass values by 20--30\% with a corresponding decrease in sail area. One such possible advance is the idea of using the APPLE RPS.  Table IV gives the results of this analysis results where the MC mass values are reduced from 97 kg to 42 kg and from 151 kg to 49 kg respectively.

Our trade study led  the following attributes of a pMC nanosat that enables formation of an MC spacecraft:\begin{enumerate}
\item Each pMC has the capability of a 6U CubeSat/nanosat. It is a 3-axis stabilized, self-contained, spacecraft able to function on its own (i.e., with ACS and communications) for a limited time. NASA JPL's MARCO CubeSat, a 6U form factor vehicle, has a mass of 13.5 kg. It carried a primary payload and operated semi-autonomously.
\item Each pMC also carries a critical part of the MC spacecraft (e.g., optical communications, extra RPS and propellant, coronagraph, primary mirror for telescope).
\item When the pMCs dock to form the MC, they share power and data.
\item The current pMC design has the shape of a round disk $\sim$1 m in diameter and $\sim$10 cm thick. The structural material is a carbon fiber composite scaffold which permits the local engineering of the CTE (e.g., James Webb Space Telescope). The assembled MC spacecraft are stacked pMC units, also in the shape of a round cylinder, but an open structure.
\item The pMCs carry a solar sail package (5 kg mass allocated), which is jettisoned at the assembly point.
\item The pMCs with solar sail package are designed to be stacked in the rideshare containers during launch. Based on mass, one ``container'' can carry multiple pMC spacecraft to form 5 MC spacecraft.
\end{enumerate}

For a MC spacecraft comprising 4 pMCs with Brayton RPS units, pMC 1 carries the primary and secondary mirrors (Fig.~\ref{fig:sc-concept}).  pMC 2 carries the science package, optical COMM and high-resolution star tracker sensors.    pMC 3 and 4 each carry additional RPS, a larger reaction wheel for attitude control, extra EP units and ``flipout'' thermal radiators.

\subsection{In-flight assembly}

The proposed architecture relies on the fact that the solar sails are not spacecraft, but a propulsion system added to the pMC. It is the pMC nanosats that provide the bus support. To allow for efficient in-flight aggregation, we choose to drop the sails prior to the in-space assembly phase.

The steps for in the in-space assembly include ejection of the sails during solar egress, at a distance between Earth and Mars orbits, and after the relative velocities of the sailcraft have been trimmed to a few cm/sec and the spatial dispersion of the pMC spacecraft ensemble is on the order of 1 km.

We studied the details of the in-space aggregation using a notional pMC spacecraft design and the power budget from the CEM tool, determining convergence statistics from $10^4$ trajectories where the pMC spacecraft dynamics was modeled in high fidelity while integrating the relevant equations of motion (i.e., central body, thrust effects).  The trajectories start from a spatial distribution comprising the pMC spacecraft within a 1 km sphere ($1\sigma$) and a velocity dispersion of 5 cm/sec ($1\sigma$) with the end goal to reach a relative distance of $< 2$~m at which proximity operation maneuvers can commence. The notional pMC spacecraft operate on the Brayton RPS with thrust values of 308 $\mu$N (at 2314 sec $I_{\rm sp}$), and 15 kg mass.

Table~\ref{tab8} presents the Monte Carlo analysis results assuming a heliocentric configuration and assembly at 1 AU. The calculation accounts for the available on-board power for thruster burns. We analyzed three thrust burn times that define the fixed-direction flight with no coasting. Using the mean values, convergence of two pMCs to a near-docking  configuration takes $\sim$6--12 hours, to a final state or position error of less than 10 cm. The amount of propellant consumed is 200--400~mg with a $\Delta v\approx$~0.3--0.6~m/s. Fewer than ten burns are required. A similar analysis has been done for assembly near Jupiter's orbit where the solar radiation pressure is down to $\sim 0.3~\mu{\rm N/m}^2$. Again, using mean values, the data shows that after sail eject it should take $\sim1.3$ days to rendezvous, consuming just 60~$\mu$g of fuel, and arriving at the final state with 0.6 m position error and a $\Delta v\approx0.12$ m/s. These results suggest that the in-space assembly need not be hastened and could take 2.4 months (travel time Earth to Jupiter) if necessary. Finally, the initial spatial dispersion of the pMC spacecraft could be as large as 10 km, but at a cost of increased fuel consumption.

\begin{table}[h!]
\caption{\label{tab8}Statistical analysis per pMC with $10^4$ samples.
}
{\small
\begin{tabular}{|>{\centering}p{0.05\textwidth}|
>{\centering}p{0.04\textwidth}|>{\centering}p{0.03\textwidth}|>{\centering}p{0.04\textwidth}|
>{\centering}p{0.04\textwidth}|>{\centering}p{0.03\textwidth}|>{\centering}p{0.04\textwidth}|
>{\centering}p{0.04\textwidth}|>{\centering}p{0.03\textwidth}|>{\centering}p{0.04\textwidth}|
>{\centering}p{0.04\textwidth}|>{\centering}p{0.03\textwidth}|>{\centering}p{0.04\textwidth}|
>{\centering}p{0.04\textwidth}|>{\centering}p{0.03\textwidth}|>{\centering}p{0.01\textwidth}l|
}\hline
Thrust burn time &
\multicolumn{3}{>{\centering}p{0.15\textwidth}|}{
Time to convergence, (days)} &
\multicolumn{3}{>{\centering}p{0.15\textwidth}|}{
Amount of propellant consumed, (g) }&
\multicolumn{3}{>{\centering}p{0.15\textwidth}|}{
Achieved final state (final position error), (m) }&
\multicolumn{3}{>{\centering}p{0.15\textwidth}|}{
Required $\Delta v$, (m/s) }&
\multicolumn{3}{>{\centering}p{0.15\textwidth}}{
\# of burn segments }&\\\cline{2-16}
~&
Mean &$3\sigma$&Worst case&
Mean  &$3\sigma$ &Worst case &
Mean  &$3\sigma$ &Worst case &
Mean  &$3\sigma$ &Worst case &
Mean &$3\sigma$&Worst case&\\\hline\hline
60\,min &0.26&0.6&1&0.22&0.58&1.0&0.08&0.23&0.35&0.34&0.88&1.6&7.1&15.2&24&\\
~&(0.11)&~&~&(0.12)&~&~&(0.05)&~&~&(0.18)&~&~&(2.7)&~&~&\\\hline
120\,min&0.33&0.6&0.8&0.26&0.56&0.8&0.07&0.22&0.5&0.4&0.85&1.2&5.2&8.57&11&\\
~&(0.09)&~&~&(0.1)&~&~&(0.05)&~&~&(0.15)&~&~&(1.13)&~&~&\\\hline
180\,min&0.5&0.95&1.1&0.41&0.95&0.95&0.08&0.26&0.3&0.61&1.42&1.4&5.39&9.29&10&\\
~&(0.15)&~&~&(0.18)&~&~&(0.06)&~&~&(0.27)&~&~&(1.3)&~&~&\\\hline
\end{tabular}
}
\end{table}

Given that the final positioning errors are $< 1$~m, onboard cameras could be used for the proximity operations for docking. The use of electromagnetic attraction to close the ``stand-off'' separation from $\sim 1.8$~m ($3\sigma$) is also possible.  A 20-turn wire about the pMC spacecraft circular shape with 0.5~A current produces a magnetic field of $\sim2.4$~mG at a distance of 1.8~m.  Applying a current with a 20~minute duty cycle, the 1.8~m separation could be closed in 1.7~hours, costing each pMC vehicle 0.17~Ah of charge.

The results of a Monte Carlo simulations shown in Table~\ref{tab8} are based on a pMC mass of 15 kg (no sails). 
The analysis of a heliocentric configuration at 1~AU, spacecraft mass 15~kg, thrust 308.64~$\mu$N, $I_{\rm sp}= 2,314.8$~s (epoch: 2045/03/21), semi-major axis: $-0.10256$, eccentricity: $1.975$, perigee radius: 0.1~AU, inclination: 0, RANN: 0, argument at periastron: 0, true anomaly: 110.83625. Thrust burn time defines flight in fixed direction, no coasting.Initial spatial dispersion 1 km ($1\sigma$), velocity dispersion 5~cm/s ($1\sigma$). The mass value used is lower than the pMC values derived using the CEM tool (20--25 kg, no sail but with mass contingency).  We foresee mass reductions with technology advances, such as 3D-printing, systems    integration, new RPS concepts, and electronics integrated into support structures that will likely close the mass differential.

 \section{Novel mission architecture}
 \label{sec:miss-arch}

The new mission architecture, enabled by in-flight aggregation, relies on the development of a reliable, mass producible spacecraft that can be propelled to high velocity by solar sail propulsion. Mass producibility can increase reliability because it enables use of specific manufacturing technologies that remove defects. Other benefits of the envisioned mission architecture include the following: i) Rideshare reduces the mission launch cost, ii) The use  of a ``parking orbit'' provides flexibility on mission scheduling; iii) Partitioning of mission functions into multiple  free-flying spacecraft that then operate in coordinated fashion enables adaptability to local failures.

Function is parsed within a cluster of spacecraft
enabling  a ``work-together'' environment that enhances adaptability, reconfigurability and redundancy all to reduce the risk of mission failure. In light of these considerations, we identify the following three key mission phases representing a mission architecture that is self-consistent and feasible.

\subsubsection{Launch phase}

Launch-related activities include all the steps from actual lift-off until a fully assembled, mission-capable spacecraft is enroute to the SGL and having achieved its desired heliocentric escape velocity and trajectory:
\begin{itemize}
\item Sailcraft are comprised of a solar propulsion system controlled by a nanosatellite-class bus, called here the proto-mission capable (pMC) spacecraft.
\item Multiple sailcraft are packed flat in containerized vehicles as secondary payloads.
\item Containerized vehicles are launched to a parking/rendezvous orbit using rideshare services.
\item Several containerized vehicles  leave the parking orbit and head toward the Sun.
\item The containerized vehicles  release sailcraft which deploy sails and begin a near circular trajectory to solar perihelion, with possible inclination change to leave ecliptic  \cite{Friedman-Garber:2014}.
\item Upon perihelion exit, sailcraft dispose extra thermal shielding (if included), reach the needed velocity and set outbound trajectory to the pre-determined SGL focal region.
\item Sailcraft fly in formation and reduce relative velocity and spatial separation.
\item While near Earth orbit, sailcraft dispose of solar sails and leaving the pMC spacecraft which aggregate and dock to form a mission capable (MC) spacecraft.
\end{itemize}

\subsubsection{Cruise phase}
\begin{itemize}
\item While flying in formation, 5--6 MC spacecraft begin a 28-year long cruise phase initially relying on optical communication and transitioning to optical navigation (e.g., via on-board astrometry).
\item At 550 AU, two MC spacecraft begin to maneuver to track the POA of the host star. Given a velocity of $>$20~AU/yr, the MC spacecraft have up to five years (550--650~AU) to find and lock on to the motion of POA of the host star (see discussion in \cite{14-Turyshev-Toth:2022-wobbles}).
\item At 650 AU, using the position location of the host star POA, the spacecraft move to find the POA of the exoplanet.
\end{itemize}

\subsubsection{Science operations phase}

\begin{itemize}
\item At 650 AU the POA of the host star becomes just a navigational ``steppingstone'' to finding the image of the exoplanet and the commencement of the science phase which could last for $\sim$10 years.
\item	The trajectory path in the science phase is non-inertial.
\item	Two approaches are considered in acquiring data on this non-inertial trajectory path. 1) The extreme case (the one presented here) where the spacecraft are in continual acceleration mode (Fig.~\ref{fig5}) and which requires more power and propellant use.  2) The more efficient power and propellant usage case where the spacecraft periodically coast using the motion of the image plane to advantage.
\end{itemize}

\section{Technologies to enable an SGL mission}
\label{sec:tech-road}

Our investigation has brought to light technologies that could impede mission success if launched today. We believe a technology roadmap can be devised to advance the readiness levels of these technologies. At the top of the list are testing methodologies that would instill confidence on component and systems reliability for the 40+ year mission. Failures are anticipated, but missions of this duration are far in few and those few (e.g., Voyager 1 \& 2) had requirements significantly different from what is needed for the SGL mission. A systems model is necessary to identify the failure modes and the available recovery approaches. Components and systems must be designed with in-situ sensors and diagnostics in mind with internal telemetry to allow near continuous monitoring of the systems health.

\subsection{Advancing solar sailing technologies}

Solar sailing propulsion offers the capability of fast solar system exit at the cost of segmenting spacecraft into low mass units. Technology advances in the development of materials that are reflective ($>$ 95\%), light (areal density $<2.5~{\rm g/m}^2$) and can be manufactured in large areas ($\sim$2000~m${}^2$) would enable propulsion speeds that could exceed 6-fold over chemical propulsion approaches. Such materials in combination with a low mass yet a stiff supporting structure, that could be folded, would be worthy advances. Recent work on graphene network structures (Aerographite, density $\sim200~\mu$g/cm${}^3$) \cite{Mecklenburg:2012} show that this cellular material is more than 4 times lighter than Ni micro lattices and lighter than reported aerogels but lack stiffness.  Perhaps impregnating these cellular structures with another material would yield a composite that would work.

 We also considered technologies \cite{Cassibry:2015} that could have profound impact on the SGL mission. Propulsion based on fusion is one such technology. A demonstration of such a vehicle would enable the transfer of metric tons of payload to the SGL in much less time than what has been shown in this analysis. A propulsion based on fission might also be a game changer \cite{12-Irvine-etal-2020}, however one has to consider the design and new protocols for increasing both safety and efficiency of launching nuclear sources into space.  The issue becomes relevant for the SGL mission as not one but numerous space systems (i.e., pMCs) will carry nuclear material (i.e., RTGs) into orbit and we anticipate multiple SGL missions to different targets. While nuclear thermal or nuclear electric propulsion would permit large mass deliveries to the SGL location, the estimated exit velocities would be $\sim$60\% slower than the 20 AU/yr base line.

\subsection{Advancing imaging technologies}

All imaging missions benefit by having larger apertures. Segmented optics are harder to implement and cost more  than monolithic. The integration time for a single SGL pixel reduces from nearly a thousand seconds for a 1-m aperture to a few hundred for a 2-m aperture. Segmented optics, launched on separate pMC spacecraft, could be assembled in space, but the assembled system must mimic the high quality ($\lambda/20$) surface of a single large optic with minimal diffraction because of the deleterious effects it could have on the coronagraph. A requirement of the primary mirror assembly is wavefront stability which depends on stiffness and thermal uniformity. Thermal stability  is achievable at the SGL location but designing for stiffness with lightweight materials is a challenge. This is because  mirror stiffness is designed for launch loads. Imagine designs if they were produced in space.  Adaptive optics in the form of deformable mirrors exist that can remove spatial frequencies associated with mirror segmenting.

The SGL imager includes a coronagraph with an extinction value requirement on the order of $10^7$. An extinction requirement 10--100 times better would increase the SNR of gleaned exoplanet photons. Traditional Lyot occulting coronagraphs are susceptible to light leakage from diffraction as that which could result from mirror segment joints. Possibly, by use of meta-optics, the high spatial frequency diffractive components could be nulled \cite{Swartzlander:2008}. In contrast to the coronagraph filtering approach, an occulting  or external starshade spacecraft approach would increase the SNR at the expense of mission complexity.

The CONOPS of the SGL mission could be simplified if optical clock technologies advance to the point where they can be low SWAP units \cite{Fortier:2019}.  Given that optical two-way time and frequency transfer has already been demonstrated in laboratories with $<$ femtometer precision will enable the harnessing of all the MC spacecraft apertures to form a single coherent imaging scenario. Synchronization of two 4 km distant optical clocks, with optical turbulence included, has been demonstrated with 225 attoseconds accuracies \cite{Deschenes:2016}.  For the SGL mission, the 4, 1-m class SGL telescopes could now be viewed as a single 4 m aperture.   This capability significantly increases the SNR and because one can add additional MC spacecraft, the possibility exists for forming even larger apertures.   The frequency comb technology which is the basis for these optical clocks have already been used in astrophysical spectrography \cite{Glenday:2015}.  Current frequency comb units have volume less than 1U and there is technology in development to get to chip scale (i.e., microcombs) devices \cite{Stephanie:2018}.

\subsection{Advancing material science}

Technology developments are required for mirror assemblies having the same properties as ULE (zero thermal expansion, Titania silicate glass) but with lower mass density. NASA's Advanced Mirror Technology Program has been developing mirrors for the next generation UVOIR telescope (4-m diameter) using a concept that fuses 3 structural core layers with 2 face sheets of ULE and Zerodur\textsuperscript{TM}.

Advancements in low outgassing materials (i.e., contamination) is another area beneficial to the SGL mission. There is need for structural materials that are high-temperature compatible ($\sim$700${}^\circ$C), low-mass (density $\sim$180 g/m${}^3$), but yet low-outgassing (total mass loss $\ll 1$\%; collected volatile condensable material $\ll$ 0.01\%).

Other advances that have basis in materials science are electronics that can operate over a wider temperature range, radiation   hardened batteries and electronics. Communications lasers with a 10-fold increase in reliability (current mean time to failure (50\%) are $\sim$19,000 hours when operating at 80${}^\circ$C and 34,000 hrs, when operating at 60${}^\circ$C) and sensor arrays that have very high dynamic range ($\sim$150 dB, currently $\sim$100 dB). Finally, materials that are multifunctional (e.g., excess structure that is converted to propellant) would be useful \cite{Dragnea:2021}.

\subsection{Advancing long-term reliability}

The length of the SGL mission poses an issue of optical sensor contamination via self-produced outgassing. These   would be volatile condensable materials. While not a showstopper, it will have to be addressed. Technology on contamination control has advanced such that key segments can be protected from contamination build up by a) proper choice of materials, b) proper venting and c) local heating. Using data from ESA's Rosetta mission\footnote{https://www.esa.int/Enabling\_Support/Operations/Rosetta}  spacecraft which was designed with contamination control in mind and had contaminant particle monitors on board, the spacecraft reached a particle density plateau of $\sim 5\times 10^{11}$  after 2500 days in orbit \cite{Liu:2021}.

If a similar contamination density is produced near after launch, then for the SGL mission, we estimate a $\sim5.3~\mu$m  of buildup over a 45-year SGL journey. This estimate is based on gas kinetic rates given a spacecraft temperature of 80 K (the coldest parts of spacecraft), an adsorbed layer sticking coefficient of 1 and an average atomic contaminant mass of 12 amu (i.e., carbon).  It does  not include UV/VUV photochemistry (for external surfaces) which could increase buildup by formation of radicals and charged species. This amount of buildup (typically of polymers) would be detrimental to the optical/imaging sensors. Rosetta implemented sensor/heating technology and proper spacecraft venting designs and an open spacecraft structure. All of that should  be implemented in the SGL spacecraft design. Additional active approaches might be necessary to maintain key sensors free of all forms of contamination. One possible solution (currently at TRL 2) is the use of ultrasonic excitation to enhance molecular diffusion on critical surfaces. A 19-fold enhancement in molecular adsorbate diffusion has been measured (4500-fold estimated) for gold clusters on silicon surfaces \cite{27-Shugaev-etal:2015}.

Beyond the recommendation for accelerated testing for reliability, this mission architecture implements a number of processes designed to enhance systems reliability, namely: i) The SGL smallsat   and its parts are intended to be mass-producible. In so doing, manufacturing technologies and methodologies can be implemented to remove defects through iteration. Moreover, mass production also reduces cost because the non- recurring cost of the design and development can be amortized over the larger number of builds. ii) Redundancy is built into the mission by having several SGL smallsats, each of which are able to do the task. Moreover, the SGL smallsat can be disassembled and reassembled in space to remove failed units. iii) The computational architecture is distributed, permitting processors from any spacecraft to take over tasks as necessary.

\subsection{Improving mission management and onboard processing}\label{sec:missman}

A mission to the SGL will require advances in spacecraft autonomy. Several factors can drive the need for enhanced    spacecraft autonomy including the need for fast reaction times, lack of a communications path to Earth, latency of communications, and limited bandwidth. To help define where autonomy might be needed and characterize the type of autonomy required, we parse the mission into several critical use cases: 
i) perihelion passage,
ii) self-assembly,
iii) search, acquisition and tracking host star and the exoplanet, and
iv) integrated vehicle health management (IVHM).
Perihelion passage is critical for establishing the correct outbound trajectory. Communication with the ground will be limited and the attitude changes might be quick. While the propulsion system (solar sail) and environmental uncertainties (solar photon flux and temperature) are unproven, autonomy at perihelion will function in a manner   very similar to traditional closed-loop guidance, navigation and control (GNC) systems, with some differences in the environment sensors and actuators.

In self-assembly the autonomous formation of an MC spacecraft via docking of pMC nanosatellites, communication to Earth should be prompt but will be constrained to low data rates.  These factors point to the need for autonomy.

We have presented the CONOPS for operations in the science phase (search, acquisition and tracking) above and this will clearly have to be autonomous.

A robust IVHM will be essential to keep the spacecraft operating over a 45-year mission lifetime. At the start of data collection at 650 AU the round-trip light time to Earth is one week, making ground resolution of time-critical faults infeasible. The spacecraft may need multiple anomaly detection techniques. IVHM will also need to decide on  a course of action to correct the anomaly, monitor results, and try other actions if the problem isn't resolved.  Additional software might be necessary that can track the ``unknown unknowns'' for reliability.  Concepts like NASA Goddard's RAISR (Research in Artificial Intelligence for spacecraft resilience\footnote{https://www.nasa.gov/feature/goddard/2021/-ai-could-speed-fault-diagnosis-in-spacecraft/}) help both on the production line and in space.  RAISR-like modules cannot work if the components are not integrated with in situ sensors that help to identify failures before mission ending scenarios arise.  The conclusions, noted above, are from a first-order architecture design for the SGL mission management software \cite{40-Neff-2021}.

The degree of autonomy also depends on the performance of the computing processors on board. Currently, there are high performance chipsets for the terrestrial market (e.g., Jetson, Corel, Myriad 2) which can operate at $10^{11}-10^{13}$  
operations/W but have TID limits of $< 100$~krad, which will not satisfy the SGL mission requirements.

However, we estimate that in a decade high performance spaceflight computing (HPSC) chiplet \cite{41-Doyle-2018,42-Shan-2022}  based processors will have similar performance with 1 MRad TID capability. HPSC aims to produce ``chiplets'' that are modular and reconfigurable, closer to ``system on a chip'' designs used in cellphones. There is an increase in single event effects (SEEs) with increase in computation speed. Devices tolerant at low speed ($< 100$~MHz) seem to have vastly increased SEEs at high speed ($>1$~GHz). Memory is also susceptible to radiation effects. This points to a need for a mixture of highly integrated circuits and lower circuit density technology. Strategies for dealing with radiation effects include 1) shielding electronics with aluminum or lead, 2) redundancy, and 3) hardening by design. Another approach involves use of wide bandgap semiconductors, e.g., gallium nitride and zinc oxide. Gallium nitride is also an attractive candidate for high temperature electronics. It is being considered for use on Venusian landers which must survive survive $\sim$460${}^\circ$C \cite{43-Eisner-etal:2021}.   An intriguing alternative is to explore electronic circuits basd on nanoscale vacuum channel transistors \cite{44-Xu-etal:2020,45-Han-2017}. These  structures are promising candidates due to their high bandwidth, high reliability, and radiation hardness.

\section{Conclusions and next steps}
\label{sec:concl}

In designing this mission some thought has been placed on what the next immediate steps should be; these include i) demonstrate propulsion control authority in sailcraft destined for a perihelion transit about the sun, ii) develop accelerated life testing for electronics, lasers, batteries and processors, and iii) develop miniaturized RPS units for distributed power.

We have presented a mission architecture for sending a spacecraft with a visible imaging payload to the SGL focal point beyond 650 AU. The proposed architecture maintains that an imaging mission to the SGL focal region is feasible.

The architecture considered here is based on the use of mass producible nanosats that get accelerated to high velocity via solar sail propulsion pass through a solar perihelion to reach a rendezvous point where the nanosats, minus  sails, self-assemble to form a MC spacecraft. We realize that this architecture fundamentally changes how   space exploration could be conducted. One can imagine small to medium scale spacecraft on fast-travelling scouting missions on quick cadence cycles which are then followed by flagship-class space vehicles. The proposed mission architecture leverages a global technology base driven by miniaturization and integration, and other technologies that are coming into fruition including composite materials based on hierarchical structures, edge computing platforms, small scale power generation and storage. These advances have had an effect on the small spacecraft industry with the development of a worldwide CubeSat and nanosat ecosystem that have continually demonstrated increasing functionality in missions (\cite{Nanosatellite:2022} and references therein).

The SGL mission is challenging. This paper presents an approach to realizing this audacious mission. The potential science return of such a mission would be unprecedented, comparable even to what would be achieved by an  actual interstellar mission, which is not achievable with present-day technology. In this report we demonstrate that, challenges notwithstanding, an SGL mission is feasible and can be accomplished using technologies that are presently available  or under active development. The anticipated discovery of life-bearing with the demonstrated feasibility of an SGL mission, should present a compelling case for pursuing this mission.  It is our   only means, in the foreseeable future, to learn details about exosolar sister planets like our home world.

\section*{Funding Sources}

The authors along with the collaborative team at NASA JPL, The Aerospace Corporation, UCLA and Xplore~Inc. acknowledge the support from the NASA Innovative Advanced Concepts Program under the 2020 NIAC Phase III grant \#80HQTR20NOA01-20NIAC\_A3, ``Direct multipixel imaging and spectroscopy of an exoplanet with a solar gravitational lens mission'' for funding this investigation. 

\section*{Acknowledgments}

The architecture study for the SGL mission was performed at The Aerospace Corporation. However, nothing is done in vacuum, and we thank Louis Friedman of The Planetary Society, Darren Garber of Xplore, Inc., Thomas Heinsheimer, Amy O'Brien, and Zigmond Lesczynski of The Aerospace Corporation for their insight, guidance and help in configuring this mission architecture, perhaps one of many, to show feasibility. This work in part was performed at the Jet Propulsion Laboratory, California Institute of Technology, under a contract with the National Aeronautics and Space Administration (80NM0018D0004). VTT acknowledges the generous support of Plamen Vasilev and other Patreon patrons.
%\textcopyright 2022. All rights reserved.

%\bibliography{arch-mission}

\end{document}